\documentclass[sigconf,nonacm]{acmart}

\copyrightyear{2024}
\acmYear{2024}
\acmConference[CCS '24]{Proceedings of the 2024 ACM SIGSAC Conference on Computer and Communications Security}{October 14--18, 2024}{Salt Lake City, UT, USA}
\acmBooktitle{Proceedings of the 2024 ACM SIGSAC Conference on Computer and Communications Security (CCS '24), October 14--18, 2024, Salt Lake City, UT, USA}
\acmDOI{10.1145/3658644.3690243}
\acmISBN{979-8-4007-0636-3/24/10}

\usepackage{enumitem}
\usepackage{fancybox}
\usepackage{tikz}

\usepackage{amsmath,amssymb}
\usepackage{balance}
\usepackage[normalem]{ulem}

\usepackage{color, colortbl}
\definecolor{Gray}{gray}{0.85}
\definecolor{verylightgray}{rgb}{.97,.97,.97}
\definecolor{mygreen}{RGB}{24,141,31}
\definecolor{myred}{RGB}{142,0,8}
\definecolor{mypurple}{RGB}{107,29,111}

\usepackage{hyperref}
\hypersetup{
    colorlinks=true,
    linkcolor=black,
    filecolor=black,
    urlcolor=black,
    citecolor=black,
}
\usepackage{booktabs, threeparttable}
\usepackage[normalem]{ulem}
\useunder{\uline}{\ul}{}
\usepackage{bbding}
\usepackage{multirow}
\usepackage{amsfonts}
\usepackage{listings}
\usepackage{tcolorbox}
\usepackage{caption,subcaption}
\usepackage{comment}
\usepackage[linesnumbered,ruled,noend]{algorithm2e}

\definecolor{verylightgray}{rgb}{.97,.97,.97}

\lstdefinelanguage{Solidity}{
	keywords=[1]{emit, anonymous, assembly, assert, balance, break, callcode, case, catch, class, constant, continue, contract, debugger, default, delegatecall, delete, do, else, event, export, external, false, finally, for, function, gas, if, implements, import, in, indexed, instanceof, interface, internal, is, length, library, log0, log1, log2, log3, log4, memory, modifier, new, payable, pragma, private, protected, public, pure, push, require, returns, revert, selfdestruct, send, storage, struct, suicide, super, switch, then, this, throw, transfer, true, try, typeof, using, value, view, while, with, addmod, ecrecover, keccak256, mulmod, ripemd160, sha256, sha3}, %
	keywordstyle=[1]\color{blue}\bfseries,
	keywords=[2]{address, bool, byte, bytes, bytes1, bytes2, bytes3, bytes4, bytes5, bytes6, bytes7, bytes8, bytes9, bytes10, bytes11, bytes12, bytes13, bytes14, bytes15, bytes16, bytes17, bytes18, bytes19, bytes20, bytes21, bytes22, bytes23, bytes24, bytes25, bytes26, bytes27, bytes28, bytes29, bytes30, bytes31, bytes32, enum, int, int8, int16, int24, int32, int40, int48, int56, int64, int72, int80, int88, int96, int104, int112, int120, int128, int136, int144, int152, int160, int168, int176, int184, int192, int200, int208, int216, int224, int232, int240, int248, int256, mapping, string, uint, uint8, uint16, uint24, uint32, uint40, uint48, uint56, uint64, uint72, uint80, uint88, uint96, uint104, uint112, uint120, uint128, uint136, uint144, uint152, uint160, uint168, uint176, uint184, uint192, uint200, uint208, uint216, uint224, uint232, uint240, uint248, uint256, var, void, ether, finney, szabo, wei, days, hours, minutes, seconds, weeks, years},	%
	keywordstyle=[2]\color{teal}\bfseries, 
	keywords=[3]{block, blockhash, coinbase, difficulty, gaslimit, number, timestamp, msg, data, gas, sender, sig, value, now, tx, gasprice, origin},	%
	keywordstyle=[3]\color{violet}\bfseries,
	keywords=[4]{onlyOwner},
	keywordstyle=[4]\color{red}\bfseries,
	identifierstyle=\color{black},
	sensitive=false,
	comment=[l]{//},
	morecomment=[s]{/*}{*/},
	commentstyle=\color{gray}\ttfamily,
	stringstyle=\color{red}\ttfamily,
	morestring=[b]',
	morestring=[b]"
}

\SetKwRepeat{Do}{do}{while}

\usepackage{wasysym}
\usepackage{pifont}
\newcommand*\circlednum[1]{\raisebox{.5pt}{\textcircled{\raisebox{-.9pt} {#1}}}}

\newcommand{\sequencer}[1]{Sequencer}
\newcommand{\executor}[1]{Executor}
\newcommand{\prover}[1]{Prover}
\newcommand{\aggregator}[1]{Aggregator}
\newcommand{\rom}[1]{ROM}
\newcommand{\validator}[1]{Validator}

\newcommand{\ignore}[1]{}

\usepackage{algorithmic}
\usepackage{graphicx}
\usepackage{textcomp}
\usepackage{xcolor}

\newcommand{\tool}{\textsc{fAmulet}}
\newcommand{\zkcode}[1]{\texttt{#1}}

\theoremstyle{definition}

\begin{document}
\title{\tool{}: Finding Finalization Failure Bugs in Polygon zkRollup}

\author{Zihao Li}
\affiliation{
\institution{The Hong Kong Polytechnic University}
\city{Hong Kong}
\country{China}
}

\author{Xinghao Peng}
\affiliation{
\institution{The Hong Kong Polytechnic University}
\city{Hong Kong}
\country{China}
}

\author{Zheyuan He}
\affiliation{
\institution{University of Electronic Science and Technology of China}
\city{Chengdu}
\country{China}
}

\author{Xiapu Luo}
\authornote{Corresponding authors}
\affiliation{
\institution{The Hong Kong Polytechnic University}
\city{Hong Kong}
\country{China}
}

\author{Ting Chen}
\authornotemark[1]
\affiliation{ 
\institution{University of Electronic Science and Technology of China}
\city{Chengdu}
\country{China}
}

\renewcommand{\shortauthors}{Zihao Li, Xinghao Peng, Zheyuan He, Xiapu Luo, and Ting Chen}

\begin{abstract}
Zero-knowledge layer 2 protocols emerge as a compelling approach to overcoming blockchain scalability issues by processing transactions through the transaction finalization process.
During this process, transactions are efficiently processed off the main chain.
Besides, both the transaction data and the zero-knowledge proofs of transaction executions are reserved on the main chain, ensuring the availability of transaction data as well as the correctness and verifiability of transaction executions.
Hence, any bugs that cause the transaction finalization failure are crucial, as they impair the usability of these protocols and the scalability of blockchains.

In this work, we conduct the first systematic study on finalization failure bugs in zero-knowledge layer 2 protocols, and define two kinds of such bugs.
Besides, we design \tool{}, the first tool to detect finalization failure bugs in Polygon zkRollup, a prominent zero-knowledge layer 2 protocol, by leveraging fuzzing testing.
To trigger finalization failure bugs effectively, we introduce a finalization behavior model to guide our transaction fuzzer to generate and mutate transactions for inducing diverse behaviors across each component (e.g., \sequencer{}) in the finalization process.
Moreover, we define bug oracles according to the distinct bug definitions to accurately detect bugs.
Through our evaluation, \tool{} can uncover twelve zero-day finalization failure bugs in Polygon zkRollup, and cover at least 20.8\% more branches than baselines. 
Furthermore, we employ \tool{} to uncover zero-day bugs and reconfirm known bugs in Scroll zkRollup and Optimism Rollup, highlighting the generality of \tool{} to be extended to other layer 2 protocols.
At the time of writing, all our uncovered zero-day bugs have been confirmed and fixed by the corresponding official teams.

\end{abstract}


%\ccsdesc[500]{Security and privacy~Distributed systems security}


\maketitle

\section{Introduction}
\label{sec_intro}
Blockchains are undergoing rapid evolution, and have gained significant recognition and attention from public communities.
To ensure their data security and integrity, blockchains employ complicated security mechanisms, e.g., distributed network and consensus protocols~\cite{bamakan2020survey}.
Nevertheless, the intricacy of these mechanisms naturally constrains blockchain performance~\cite{bamakan2020survey}. For example, Ethereum can only process tens of transactions per second~\cite{paavolainen2020security}.

To mitigate blockchain scalability issues (e.g., increasing transaction throughput), identified as a critical blockchain performance bottleneck~\cite{kim2018survey}, 
zero-knowledge layer 2 (L2) protocols have been proposed as a promising approach to enhance the scalability of Ethereum~\cite{vcapko2022state}.
Their core idea involves executing transactions on L2 blockchains, which forgoes the burdensome security features to improve transaction processing speed~\cite{vcapko2022state}.
The transaction data from L2 blockchain is then posted to L1 blockchain (e.g., Ethereum) as payload data of L1 transactions~\cite{vcapko2022state}.  
Since the transaction data from L2 blockchain is finalized on L1 blockchain, the security of these L2 transactions is assured by the security mechanisms of L1 blockchain.
Additionally, these protocols utilize zero-knowledge (zk) proof techniques~\cite{sun2021survey,miyahara2024zkp,yang2023zero,ma2024research,li2024trusted} to generate zk proofs, which can verify the correctness of transaction execution results executed on L2 blockchain~\cite{vcapko2022state}.
By posting these zk proofs to L1 blockchain, the correctness of the transaction execution on L2 blockchain can be efficiently validated through a single L1 transaction~\cite{polygon_zkevm_doc}.

Since L2 transactions occur and are processed off L1 blockchain, their data availability and security depend on the successful completion of the transaction finalization process.
This process involves processing transactions on L2 blockchain and reserving both the transactions and their execution proofs on L1 blockchain~\cite{koegl2023attacks}.
Upon completing this process, L2 transactions are finalized on both L1 and L2 blockchains, and the correctness and verifiability of their execution results are guaranteed~\cite{polygon_zkevm_doc}.
Hence, any bugs in these protocols that disrupt the transaction finalization process (termed as \underline{finalization failure bugs}) can impair the usability of these protocols and the scalability of L1 blockchains~\cite{koegl2023attacks}, and even result in financial losses for users~\cite{polygon_incident,koegl2023attacks,medium_offchainlabs_rollup_delay}.
For example, cross-chain bridges that respond to the L2 transactions finalized on L1 blockchain will malfunction when the transaction finalization process halts, posing a risk of financial loss to users~\cite{koegl2023attacks,medium_offchainlabs_rollup_delay}.

Fuzzing techniques have been proven as a promising technique in effectively uncovering various types of bugs within modern software systems~\cite{schloegel2024sok}.
Through leveraging fuzzing techniques, testing tools such as  Fluffy~\cite{yang2021finding}, LOKI~\cite{ma2023loki}, Tyr~\cite{chen2023tyr}, and EVMFuzzer~\cite{fu2019evmfuzzer} have been applied to the blockchain domain, where they have successfully detected critical bugs within blockchain systems.
However, to effectively find finalization failure bugs in zero-knowledge L2 protocols,
there are three major challenges for these tools.

\noindent
\textit{\textbf{Challenge 1: Finalization failure bugs are hard to trigger effectively.}}
The transaction finalization process involves complicated interactions among various L1 and L2 components. However, as current zero-knowledge L2 protocols typically operate in a centralized manner by a permissioned entity, 
the testing tool is unable to directly manipulate each component like existing tools~\cite{chen2023tyr,ma2023loki}. 
Consequently, the testing tool is restricted to conducting tests by only submitting L2 transactions to find finalization failure bugs.
This scenario necessitates relying solely on mutating transaction inputs to explore the behaviors of each component within the intricate logic path, which poses a significant challenge.

\noindent
\textit{\textbf{Challenge 2: Generated transactions triggering error logic are preemptively discarded.}}
Current zero-knowledge L2 protocols incorporate a transaction pool that includes a pre-execution phase.
This phase aims to preemptively discard as many transactions as possible triggering execution errors before they are added to the transaction pool. 
Hence, even if the testing tool successfully generates transactions that induce error logic, these transactions are likely to be preemptively discarded prior to processing them on L2 blockchain, thereby reducing the efficiency of bug detection.

\noindent
\textit{\textbf{Challenge 3: It is hard to precisely detect and locate finalization failure bugs.}}
The interactions between L1 and L2 components are dynamic and complicated.
Hence, detecting and locating finalization failure bugs requires precise oracles tailored for these bugs.

We propose \tool{}, the first testing tool for detecting finalization failure bugs in Polygon zkRollup~\cite{polygon_zkevm_doc}, a prominent zero-knowledge L2 protocol, by leveraging fuzzing techniques.
To address \textbf{C1}, \tool{} employs a finalization behavior model that records runtime data (e.g.,  the logic branches executed) to depict runtime behaviors of each component involved in the finalization process. This model augments \tool{} to continuously and efficiently mutate inputs (i.e., transactions) to explore diverse behaviors within the finalization process, aiming to expose the finalization failure bugs.
Furthermore, \tool{} integrates a behavior guided transaction fuzzer equipped with several mutation strategies to generate and mutate transactions with varied execution contexts (e.g., stack and memory operations), enhancing the potential to trigger finalization failure bugs.
To address \textbf{C2}, we integrate a logic injection technique to conceal code logic within test transactions during the pre-execution phases, thereby ensuring that the transactions generated to trigger error logic will be processed on L2 blockchain.
Moreover, to address \textbf{C3}, we design bug oracles tailored to detect two kinds of finalization failure bugs, derived from the two stages of finalization process. Besides, \tool{} leverages the collected runtime information from the finalization behavior model to locate finalization failure bugs and ascertain their root causes.

We implement \tool{} and conduct extensive experiments to evaluate its effectiveness in terms of detecting finalization failure bugs.
Our experimental results demonstrate that \tool{} successfully identifies twelve zero-day finalization failure bugs on Polygon zkRollup that can disrupt the finalization process.
Moreover, by comparing with two baseline tools established by us, Fluffy-F and Fuzzer-Cover, \tool{} can cover Fluffy-F by 20.8\% and Fuzzer-Cover by 31.3\% in branch coverage, respectively.
Furthermore, we employ \tool{} to uncover a zero-day finalization failure bug in Scroll zkRollup, and reconfirm two known finalization failure bugs from audit reports in Optimism Rollup, highlighting the generality of \tool{} to be extended to other layer 2 protocols.

We derive four insights on why finalization failure bugs occur and how they can be fixed, originating from our investigation of bug root causes and discussions with the official teams.
We disclose all identified bugs to Polygon zkRollup team and Scroll zkRollup team via the Immunefi platform~\cite{immunefi}.
At the time of writing, all these bugs have been confirmed and fixed by the two official teams, and they have awarded us corresponding bug bounties to acknowledge and highlight the practical significance of our findings.

\noindent
\textbf{Contributions.}
We summarize our contributions as follows:

\begin{itemize}[leftmargin=*,topsep=1pt]
    \item We conduct the first systematic study on finalization failure bugs in zero-knowledge L2 protocols, and define two kinds of bugs based on the two stages of the transaction finalization process.
    \item We design and implement \tool{}, the first tool for detecting finalization failure bugs in Polygon zkRollup. We introduce a finalization behavior model to guide our transaction fuzzer in exploring diverse behaviors within the finalization process. Besides, we define bug oracles according to the distinct definitions of finalization failure bugs to accurately detect them.

    \item We conduct experiments to evaluate the effectiveness of \tool{} in detecting finalization failure bugs. \tool{} successfully detects twelve serious zero-day bugs within Polygon zkRollup, which have been confirmed and repaired. Besides, compared with baselines, \tool{} can cover at least 20.8\% more branches.
    \item We derive new insights and understandings on why finalization failure bugs occur and how they can be fixed. Our insights will not only enhance the security of Polygon zkRollup but also provide valuable guidance for other zero-knowledge L2 protocols.
\end{itemize}

\noindent We refer readers to~\cite{appendixpaper} for our full paper version with the appendix.

\section{Background}
\label{sec_background}

\subsection{Polygon zkRollup overview}
\label{sec_back_archi}

\begin{figure}[b]
	\centering
	\includegraphics[width=0.94\linewidth]{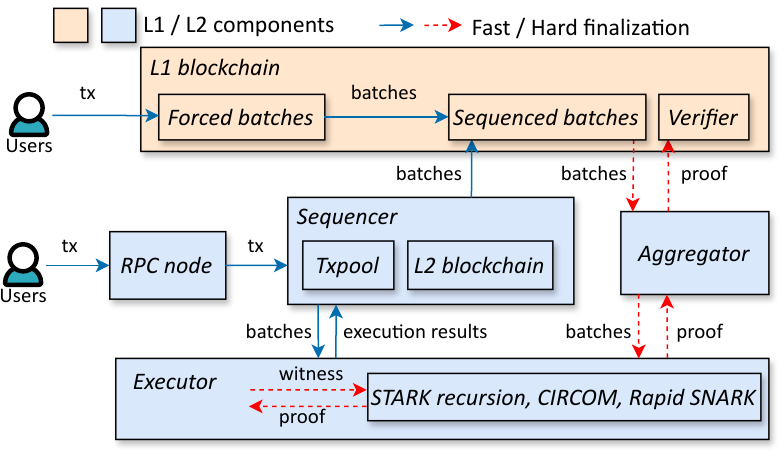}
	\caption{Architecture of Polygon zkRollup.}
	\label{fig_arch_zkevm}
\end{figure}

Polygon zkRollup is a zero-knowledge L2 protocol designed to boost the throughput of Ethereum transactions~\cite{polygon_zkevm_doc}.
We depict the architecture of Polygon zkRollup in Fig.~\ref{fig_arch_zkevm}. 
Please note, although various zero-knowledge L2 protocols~\cite{scroll} may have different implementations, they feature similar components in their finalization processes akin to those in Polygon zkRollup.
In Fig.~\ref{fig_arch_zkevm}, we abstract the architecture of Polygon zkRollup with its finalization process in a unified way, facilitating the generality of our study.

As shown in Fig.~\ref{fig_arch_zkevm}, upon receiving transactions from L2 users, Polygon zkRollup will process these transactions within its L2 blockchain, group the processed transactions into batches, and then submit these batches as payload data in Ethereum (L1 blockchain) transactions to the L1 blockchain~\cite{polygon_zkevm_doc}.
Additionally, Polygon zkRollup generates zero-knowledge proofs for the processed L2 transactions to ensure that i) the L2 transactions are executed correctly, and ii) the updated L2 blockchain state is accurate.
Consequently, transactions on L2 blockchain are executed off the L1 blockchain while still benefiting from the underlying blockchain’s security.
In the following, we will detail how each component contributes to the functionalities of Polygon zkRollup.

\noindent
\textbf{RPC node.}
RPC node provides HTTP interfaces that allow users to interact with Polygon zkRollup.
For example, by initializing requests to an RPC node, users can submit transactions to L2 blockchain, and query the current state of L2 blockchain.

\noindent
\textbf{Txpool.}
User transactions received from the RPC node are initially placed into a local transaction pool (txpool)~\cite{li2021deter}.
Txpool will pre-execute these transactions, discarding any that encounter execution errors~\cite{polygon_zkevm_doc}.
We elaborate on what kinds of execution errors will be discarded during the pre-execution phase in Appendix A.
After being included in txpool, transactions are sorted based on their efficiency~\cite{polygon_zkevm_doc}.
Notably, akin to the gas price~\cite{wood2014ethereum}, efficiency refers to the amount of funds users are prepared to expend to have their transactions preferentially included in L2 blockchain~\cite{polygon_zkevm_doc}.

\noindent
\textbf{Sequencer.}
By loading user transactions from the txpool, \sequencer{} is responsible for ordering these transactions based on their efficiency, assembling them into batches, and subsequently submitting these batches to L1 blockchain.
Once submitted, the batches are sequenced in the predetermined order~\cite{polygon_zkevm_doc}.
Moreover, \sequencer{} will forward batches to \executor{} for transaction processing, and update the local L2 blockchain state based on the execution results of the transactions in each batch returned by \executor{}.

\noindent
\textbf{Executor.}
\executor{} integrates a virtual machine, \rom{}, for executing transactions, which functions equivalently to Ethereum Virtual Machine (EVM)~\cite{wood2014ethereum}.
The transaction execution in \rom{} is measured in execution steps (Counters), akin to the gas for Ethereum transactions~\cite{wood2014ethereum}, which are used to limit the resources expended in generating the corresponding zk proof for the transaction execution~\cite{polygon_zkevm_doc}. 
\rom{} is developed using two novel languages, namely zkASM and PIL~\cite{polygon_zkevm_doc}, to enable the generation of zero-knowledge proof for transaction execution.
When interpreting transaction execution in \rom{}, \executor{} will generate a witness for each transaction execution, including transaction execution traces.
Besides, during generating witnesses for transactions, \executor{} checks whether these traces satisfy predefined constraints.
The constraints dictate the rules each operation within a transaction must adhere to, ensuring that modifications to L2 blockchain state are both valid and verifiable through zero-knowledge proofs (Appendix A).
After processing transactions received from \sequencer{}, \executor{} returns the transaction execution results to \sequencer{} to update the local L2 blockchain state.
In the process of proof generation for transaction execution, \executor{} will transmit the generated witness to the cryptographic proving backend for further proof generation.

The cryptographic proving backend within \executor{} comprises several cryptographic tools (e.g., STARK recursion, CIRCOM, and zk-SNARK), which work in concert to generate valid zk proofs for the transaction execution witness.
Specifically, leveraging the Fast Reed-Solomon Interactive Oracle Proofs of Proximity~\cite{ben2018fast}, the STARK recursion~\cite{ben2019scalable} compresses multiple proofs into a recursive zk-STARK proof to scale and accelerate the proof generation process.
The CIRCOM library~\cite{belles2022circom} then constructs an arithmetic circuit based on the zk-STARK proof from the STARK recursion, generating valid input, intermediate, and output values for the circuit.
This information is subsequently processed by Rapid SNARK~\cite{polygon_zkevm_doc} to generate a valid zk-SNARK proof~\cite{petkus2019and}. Ultimately, the generated zk proofs serve to verify the correctness of the transaction execution and the updated L2 blockchain state.

\noindent
\textbf{Aggregator.} 
\aggregator{} monitors and gathers batches submitted by \sequencer{} from L1 blockchain.
By interacting with \executor{} using the gathered batches, \aggregator{} is able to obtain zero-knowledge proofs for the batches.
These proofs can be used to validate the correctness of the transactions within the batches.
Subsequently, \aggregator{} publishes the zero-knowledge proofs onto L1 blockchain, allowing any party to verify the proofs' validity by submitting corresponding transactions to the L1 blockchain.

\noindent
\textbf{Forced batches.}
Forced batches are implemented to maintain the liveness of Polygon zkRollup in scenarios where L2 components become entirely unresponsive or behave maliciously~\cite{polygon_zkevm_doc,forced_batches}. 
For example, even if the L2 components such as \sequencer{} stop working, users can still directly submit L2 transactions to Forced batches on L1 blockchain. The Forced batches ensure the inclusion of batches containing the L2 transactions in the sequenced batches, thereby preserving the liveness of Polygon zkRollup.
Once these batches are forcibly included in the sequenced batches on L1 blockchain, \aggregator{} collects these batches, facilitates the generation of corresponding zero-knowledge proofs, and subsequently publishes the proofs for the forcibly included batches.

\subsection{Finalization process in Polygon zkRollup}
\label{sec_back_final}
The transaction finalization process is a core functionality of Polygon zkRollup to enable the scalability of Ethereum~\cite{polygon_zkevm_doc}.
We depict the whole process in Fig.~\ref{fig_arch_zkevm}.
In this process, L2 transactions are first processed within L2 blockchain, then grouped into batches, and subsequently submitted to L1 blockchain. 
Besides, the zk proofs validating the correctness of their execution are generated and published on L1 blockchain.
Upon completing this process, the L2 transactions are finalized on both L1 and L2 blockchains, ensuring that their execution results are both correct and verifiable.

Highlighted by the \textcolor[RGB]{0,100,162}{blue} and \textcolor[RGB]{254,0,19}{red} lines in Fig.~\ref{fig_arch_zkevm}, the transaction finalization process is divided into two stages, i.e., fast finalization and hard finalization.
The two stages correspond to the two levels of transaction finalized states provided by Polygon zkRollup~\cite{polygon_zkevm_doc}.

\noindent
$\bullet$
\textbf{Fast finalization.}
Transactions reach a fast-finalized state when they have been processed and included in a block on L2 blockchain, with corresponding batches containing them prepared for posting to L1 blockchain.
The fast finalization process of L2 transactions begins when users submit their transactions to Polygon zkRollup via the RPC node, and concludes once the transactions reach the fast-finalized state~\cite{polygon_zkevm_doc}.
Third-party applications on Polygon zkRollup typically react to the execution results of transactions after their fast finalization process is complete, as this approach enhances their efficiency and response time~\cite{polygon_dapps}.
For example, a wallet will display the execution results to users who initialize the transactions once the corresponding soft finalization process has concluded.

\noindent
$\bullet$
\textbf{Hard finalization.}
Transactions reach a hard-finalized state when the zk proofs of the transactions have been generated and published on L1 blockchain.
The hard finalization process of L2 transactions begins when batches containing these transactions are sequenced on L1 blockchain, and concludes once the transactions reach the hard-finalized state.
Please note that the execution results of a transaction are deemed final and immutable only when the transaction reaches the hard-finalized state~\cite{polygon_zkevm_doc}.
However, if a transaction is in the soft-finalized state, the transaction may still be rolled back on L2 blockchain under special circumstances~\cite{polygon_zkevm_doc}, such as L1 blockchain reorganization~\cite{schwarz2022three}.
Therefore, in scenarios demanding the highest level of security, e.g., cross-chain transfers, third-party applications typically respond only to the execution results of transactions that have reached the hard-finalized state~\cite{polygon_dapps}.

\subsection{Finalization process in Optimistic L2s}
\label{sec_bac_opl2}

Optimistic L2 protocols represent another  category of layer 2 protocols aimed at enhancing the scalability of Ethereum~\cite{sguanci2021layer}.
Unlike zero-knowledge L2 protocols, which employ zero-knowledge proof techniques to ensure the validity of transaction executions off the main chain, optimistic L2 protocols employ interactive fraud proof techniques~\cite{kalodner2018arbitrum} to redress incorrect state transitions submitted on L1 blockchain, thereby ensuring the correctness of transaction executions off the main chain~\cite{kalodner2018arbitrum}.
The transaction finalization process in optimistic L2 protocols also comprises fast and hard finalization processes, similar to those adopted in zero-knowledge L2 protocols.

In the fast finalization process, 
optimistic L2 protocols also employ: i) RPC node to receive transactions from users, ii) Txpool to reserve transactions passed the pre-execution phase, and iii) \sequencer{} to order and assemble transactions into batches, submit batches to L1 blockchain, and maintain a local L2 blockchain with updating state transitions.
The main difference in the fast finalization process between optimistic and zero-knowledge L2 protocols lies in \executor{}, i.e., the component for executing transactions and generating state transitions.
Zero-knowledge L2 protocols typically build \executor{} from scratch using zk-friendly languages and primitives (e.g., Poseidon hash~\cite{grassi2021poseidon}) to facilitate efficiently generating zk proofs for state transitions. 
In contrast, optimistic L2 protocols, without specific requirements, commonly adopt the official repository of the L1 blockchain like Ethereum, to build \executor{}~\cite{Optimism}.

In the hard finalization process, optimistic L2 protocols introduce \validator{} instead of \aggregator{} for conducting interactive fraud proofs for batches posted on L1 blockchain~\cite{kalodner2018arbitrum}.
Specifically, 
\validator{} in optimistic L2 protocols is responsible for challenging the correctness of L2 transactions submitted on L1 blockchain by initiating challenges against the validity of the transactions and their execution results~\cite{kalodner2018arbitrum}.
After the challenge period has elapsed or if the challenges do not successfully establish, the corresponding transactions reach the hard-finalized state.
In contrast, \aggregator{} in zero-knowledge L2 protocols is responsible for establishing the validity of L2 transactions submitted on L1 blockchain by generating corresponding zk proofs~\cite{polygon_zkevm_doc}. 
Only upon the generation of the zk proofs can the transactions reach the hard-finalized state~\cite{polygon_zkevm_doc}.

\section{Problem definition}
\label{sec_system_model}

\begin{figure*}[t]
	\centering
	\includegraphics[width=0.82\linewidth]{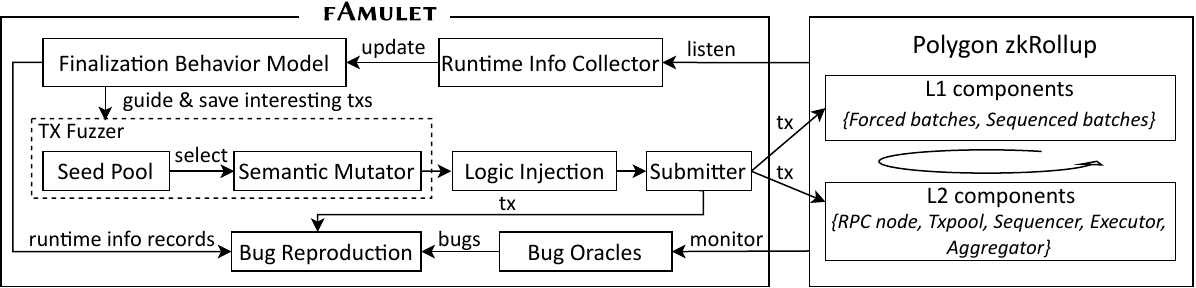}
	\caption{An overview of \tool{}. \circlednum{1} In the transaction fuzzer, \tool{} first selects test transactions from seed pool and mutates their execution context. \circlednum{2} \tool{} then disguises the mutated transactions to bypass the pre-execution checks. \circlednum{3} \tool{} submits test transactions to our Polygon zkRollup testnet for processing. \circlednum{4} \tool{} collects runtime information to update the finalization behavior model for guiding future seed selection and transaction mutation. \circlednum{5} Bug oracles monitor the testnet in real-time to detect finalization failure bugs, and \tool{} reproduces the identified bugs to facilitate the derivation of their root causes. \tool{} iterates through these five steps until its termination.
    }
	\label{fig_tool_overview}
\end{figure*}

\noindent
\textbf{System model.}
We utilize the real-world deployment environment of Polygon zkRollup as our system model to ensure the practicality and reliability of our findings.
Specifically, the L2 components of Polygon zkRollup are managed in a centralized manner, and controlled by a permissioned entity.
Users are limited to interacting with Polygon zkRollup by submitting L2 transactions through its RPC node. %
Besides, an L1 blockchain, e.g., Ethereum, operates as the underlying layer of Polygon zkRollup, storing the submitted batches and zk proofs for each batch from Polygon zkRollup.
The L1 blockchain supports the Forced batch feature (\S\ref{sec_back_archi}) of Polygon zkRollup, ensuring that Polygon zkRollup's liveness is maintained.
Furthermore, we assume the networks of both L1 and L2 components to be reliable and their communication stable. %
During our testing, all L1 and L2 components remain unaffected by external factors, including blockchain reorganization, power outages, and network intrusion. 
Moreover, external attack vectors originating from L1 blockchain, such as network partitioning, node partitioning, and byzantine attacks, are not considered in our testing.

In our system model, we define a Polygon zkRollup system as $\phi = \{TX, \mathcal{B}_{2}\}$, and an L1 blockchain as $\psi = \{\mathcal{B}_{1}, \mathcal{P}\}$.
The finite set $TX = \{tx_1, tx_2,..., tx_n\}$ presents the transaction pool within the Polygon zkRollup system.
Besides, the finite set $\mathcal{B}_{2} = \{B^1_2, B^2_2,..., B^m_2 \}$ refers to the batches generated and processed on the L2 blockchain within the Polygon zkRollup system, where each batch $B^i_2$ contains a set of confirmed L2 transactions.
Concurrently, the finite set $\mathcal{B}_{1} = \{B^1_1, B^2_1,..., B^n_1 \}$ presents the batches sequenced on the L1 blockchain.
Furthermore, the finite set $\mathcal{P} = \{p^1, p^2,..., p^m \}$ refers to the zk proofs submitted to the L1 blockchain for the sequenced batches, where each $p^i$ refers to a valid zk proof corresponding to a batch $B^i_1$ sequenced on L1 blockchain.

\noindent
\textbf{Definition 1} (Fast Finalization Failure Bugs).
Fast finalization failure (FFF) bugs violate the liveness property of the fast finalization process. Specifically,
for transactions that are included in the transaction pool of Polygon zkRollup, the batches containing them cannot be generated and processed on the L2 blockchain, and these transactions cannot reach the fast-finalized state.  
FFF bugs occur when the L2 components involved in the fast finalization process (e.g., \sequencer{}) either terminate to exit erroneously or are stuck in the fast finalization process.
Formally, FFF bugs are triggered when $\forall tx_i \in TX, \lnot(\lozenge (tx_i \in B^j_2))$, which means that for every L2 transaction in the transaction pool, there is no batch processed on L2 blockchain that will eventually include it.

\noindent
\textbf{Definition 2} (Hard Finalization Failure Bugs).
Hard finalization failure (HFF) bugs violate the liveness property of the hard finalization process. Specifically,
for batches that are sequenced on the L1 blockchain, the zk proofs to verify the correctness of these batches cannot be produced, and the L2 transactions within these batches cannot reach the hard-finalized state.
HFF bugs occur when the L2 components involved in the hard finalization process (e.g., \executor{}) either terminate to exit erroneously or are stuck in the hard finalization process.
Formally, HFF bugs are triggered when $\forall B^i_1 \in \mathcal{B}_{1}, (i > j), \lnot(\lozenge p^i)$, which means that for every sequenced batch after a specific batch ($B^j_1$), there is no a corresponding zk proof that is eventually be generated and submitted to the L1 blockchain.

\section{Overview}

\tool{} is designed to detect both fast and hard finalization failure bugs in Polygon zkRollup.
To expose finalization failure bugs, \tool{} employs a behavior guided transaction fuzzer to iteratively generate and mutate transactions as inputs to induce diverse behaviors across each component within  transaction finalization process.
By modeling the runtime behaviors of all involved components during the finalization process, \tool{} covers the search space for finding finalization failure bugs.
\tool{} utilizes this model to guide the transaction generation and mutation in next rounds to enhance the potential to trigger finalization failure bugs.
Furthermore, employing bug oracles that monitor Polygon zkRollup in real-time, \tool{} detects and locates finalization failure bugs and utilizes the context information and historical transactions to reproduce bugs, thereby facilitating the derivation of their root causes.

Fig.~\ref{fig_tool_overview} illustrates the overview of \tool{}.
\circlednum{1} Initially, the behavior guided transaction fuzzer selects transactions from the seed pool, which includes previously executed test transactions. The transaction fuzzer then mutates the execution context of test transactions with several mutation strategies. \circlednum{2} \tool{} employs a logic injection technique to disguise mutated test transactions, ensuring they are not prematurely discarded by the transaction pool of Polygon zkRollup during the pre-execution phase. \circlednum{3} After signing test transactions, \tool{} submits them to our Polygon zkRollup testnet for processing, either through the L2 RPC node or L1 Forced batches. \circlednum{4} \tool{} collects runtime execution information from each component involved in the finalization process to update the finalization behavior model for guiding the seed selection and transaction mutation in the next iterations of the transaction fuzzer. Besides, \tool{} also saves the new test transactions in the seed pool if they trigger new behaviors. \circlednum{5} In the meanwhile, bug oracles monitor our Polygon zkRollup testnet in real-time to detect and locate finalization failure bugs. For each identified bug, \tool{} utilizes the collected execution information and historical test transactions to reproduce the bug, thereby facilitating the derivation of its root cause. 
Consequently, \tool{} iterates through the five steps (from \circlednum{1} to \circlednum{5}) until its termination.

\begin{table}[t]
\caption{Behavior data of each L2 component}
\resizebox{0.72\linewidth}{!}{
\begin{tabular}{lc}
\toprule
\textbf{L2 component} & \textbf{Internal state} \\ \midrule
RPC node     &   $Signer$  \\
Txpool       &   $PoolTX$, $PoolTX_{Res}$     \\
Sequencer    &   $B$, $SeqTX_{Inv}$     \\
Executor     &   $Stack$, $Mem$, $Storage$, $Counter$    \\
Aggregator   &   $ForceBatch$, $SeqBatch$ \\ \bottomrule
\end{tabular}
}

\label{tab_internal_state}

\end{table}

\section{Design}
\label{sec_design}

We describe the core design of \tool{}, focusing on the finalization behavior model (\S\ref{sec_behavior_model}), the transaction fuzzer (\S\ref{sec_fuzzing}), the logic injection technique (\S\ref{sec_logic_injection}), and bug oracles and reproduction (\S\ref{sec_bug_analysis}).

\subsection{Finalization behavior model}
\label{sec_behavior_model}

The finalization process involves intricate interactions among various L1 and L2 components, ultimately aiming to process transactions and reserve both their transaction data and zk proofs of their executions on L1 blockchain.
Hence,
to efficiently trigger finalization failure bugs within the deep interaction logic, it is crucial to expose diverse runtime behaviors of each component.
We achieve this by employing a finalization behavior model that captures and reflects the executed behaviors, which guides the subsequent transaction fuzzer to explore different behaviors.

In our model, the finalization behaviors are characterized by two kinds of behavior data, i.e., the overall behavior data and the behavior data of each component.
The overall behavior data, denoted as $(Cover, STX_{RPC}, STX_{Force}, FTX_{Fast}, FTX_{Hard})$, indicates the global state of the finalization process, including: i) $Cover$, the set of coverage information detailing the branches covered by all components, ii) $STX_{RPC}$ and $STX_{Force}$, the transactions are sent to Polygon zkRollup via the RPC node and the Forced batches, respectively, and iii) $FTX_{Fast}$ and $FTX_{Hard}$, the transactions that reach the fast-finalized and hard-finalized states, respectively.

The behavior data of each component captures the internal state of each component during the finalization process. As we exclude external attack vectors originating from L1 blockchain (\S\ref{sec_system_model}), our focus is specifically on the behavior data of each L2 component. We list their behavior data in Table~\ref{tab_internal_state} and detail them as follows.

\noindent
\textit{-- RPC node.}
RPC node is the entry point for users to submit transactions.
Since we record the transactions received by RPC node in the overall behavior data, we focus on the signer accounts associated with these transactions, which are denoted as $Signer$.

\noindent
\textit{-- Txpool.}
Txpool stores transactions that have not yet been processed by \sequencer{}, referred to as $PoolTX$.
Please note that Txpool can include transactions returned by \sequencer{} for reservation within Txpool.
To better capture behaviors within Txpool, our model also includes these transactions, which are denoted as $PoolTX_{Res}$.

\noindent
\textit{-- Sequencer.}
After processing transactions, \sequencer{} assembles them into batches for posting to L1 blockchain.
We denote these batches as $B$, where each of them contains a sequence of transactions.
If errors are encountered during transaction processing, \sequencer{} can handle these errors and discard corresponding transactions from Txpool.
These discarded transactions are denoted as $SeqTX_{Inv}$. It is worth noting that although transactions in $SeqTX_{Inv}$ trigger errors, these errors are handled by \sequencer{} and do not impact the finalization process.

\noindent
\textit{-- Executor.}
\executor{} is responsible for executing transactions, and firstly generating the witness for transaction executions during the hard finalization process.
Meanwhile,
the cryptographic proving backend within \executor{} takes in the witness, and produces corresponding zk proofs for the transactions.
Although the witness data provides a detailed representation of transaction executions, using them to calculate the differences in the behaviors of transaction execution and zk proof generation is challenging and inefficient for two reasons.
First, the witness data consists of tables filled with binary data, which are inherently difficult to compare. Second, the substantial size of the witness data (e.g., exceeding 10 MB for a single transaction execution~\cite{polygon_zkrollup_witness}), hinders efficient comparison.
To effectively and efficiently model the internal behaviors within \executor{}, we extract four key metrics representing the transaction executions. These metrics include: i) $Stack$, the maximal stack regions utilized in each transaction, ii) $Mem$, the maximal memory regions utilized in each transaction, iii) $Storage$, the storage locations accessed during each transaction, and iv) $Counter$, the number of execution steps taken for each transaction.

\noindent
\textit{-- Aggregator.}
Aggregator monitors batches sequenced on L1 blockchain and downloads them for the process of generating zk proofs.
These batches can originate from two sources, i.e., i) the batches from Forced batches, and ii) the batches generated and submitted by \sequencer{}, denoted as $ForceBatch$ and $SeqBatch$, respectively.

\noindent
\textbf{The rationale for determining internal states.}
The selection and collection of internal states for different L2 components are inspired by how they are engaged in the finalization process. 
Otherwise, if partial internal states have already been collected in the overall behavior data, we will skip the recorded ones and focus on the remaining internal states.

\noindent
\textbf{Recognizing new execution behaviors.}
Upon monitoring the execution information of each component in real-time as transactions are processed, the finalization behavior model will be updated accordingly.
After processing a transaction $TX_{j}$, the model recognizes new behaviors if, upon updating, the overall behavior data or the behavior data of any component differs from the previously recorded behavior data in either of these two categories:

The overall behavior data of a new transaction differs from the previously recorded overall behavior data if new coverage information is included, i.e., $Cover_{j} \notin Cover$, or if new transactions reach the fast-finalized or hard-finalized states, i.e., $FTX_{Fast_{j}} \supset FTX_{Fast}$ or $FTX_{Hard_{j}} \supset FTX_{Hard}$, respectively.

The behavior data of each component during processing a new transaction differs from the previously recorded behavior data of components if new internal states of any component, compared to the previously recorded internal states, are included:

\noindent
$\bullet$
i) RPC node. There are new signer accounts $Signer_{j}$ for signing transactions, i.e., $Signer_{j} \supset Signer$. 

\noindent
$\bullet$
ii) Txpool. The transactions in transaction pool have changed, i.e., $PoolTX_{j} \neq PoolTX$, or
the transactions in transaction pool returned by \sequencer{} have changed, i.e., $PoolTX_{Res_{j}} \neq PoolTX_{Res}$. 

\noindent
$\bullet$
iii) Sequencer. There are new batches generated by \sequencer{}, i.e., $B_{j} \supset B$, or there are new transactions discarded by \sequencer{}, i.e., $SeqTX_{Inv_{j}} \supset SeqTX_{Inv}$. 

\noindent
$\bullet$
iv) Executor. There are new maximal stack regions utilized, i.e., $Stack_{j} \notin Stack$, new maximal memory regions used, i.e., $Mem_{j} \notin Mem$, new storage locations accessed, i.e., $Storage_{j} \notin Storage$, or new execution steps taken, i.e., $Counter_{j} \notin Counter$.

\noindent
$\bullet$
v) Aggregator. New batches are downloaded from the batches originating from Forced batches, i.e., $ForceBatch_{j} \supset ForceBatch$, or new batches are downloaded from batches generated and submitted by \sequencer{}, i.e., $SeqBatch_{j} \supset SeqBatch$.

\subsection{Behavior guided transaction fuzzing}
\label{sec_fuzzing}

As mentioned in~\S\ref{sec_system_model}, the runtime behaviors of components in finalization process are driven by the submitted test transactions. 
To explore diverse behaviors of components for triggering finalization failure bugs, we employ a behavior guided transaction fuzzer that continuously generates and mutates varied test transactions with the guidance of the finalization behavior model.

Compared to studies~\cite{yang2021finding} that employ multi-transaction fuzzing to find bugs in blockchain systems by generating multiple transactions invoking the same contract, we focus on single-transaction fuzzing to trigger the finalization failure bugs. This approach minimizes the inter-effects of transaction executions across different rounds, thereby enabling a more intuitive observation of the impact that each transaction has on different components within the finalization process. Therefore, the finalization behavior model can more effectively guide the transaction fuzzing to explore diverse behaviors for triggering finalization failure bugs.

Algorithm~\ref{alg_fuzzing} presents the process of behavior guided transaction fuzzing.
The fuzzer takes in initial transaction seeds and Polygon zkRollup testnet, launching the testnet (Line 2) before initiating the transaction fuzzing.
In each iteration, the fuzzer randomly selects a transaction from initial seed pools (Line 4), and mutates it for processing on testnet (Line 5).
We detail the transaction mutation later in this section.
The mutated transactions are then disguised by logic injection technique (Line 6) to ensure that it is not preemptively discarded by Txpool (cf. details of logic injection in \S\ref{sec_logic_injection}).
These disguised transactions are then submitted to Polygon zkRollup testnet via either the RPC node or Forced batches (Lines 7-10), and the finalization behavior model is updated according to real-time feedback from monitoring the testnet (Line 11).
Concurrently, the fuzzer employs bug oracles to detect any new finalization failure bugs (Line 12) and records these identified bugs (Line 13).
The detailed process of bug detection is introduced in~\S\ref{sec_bug_analysis}.
If a test transaction triggers new execution behaviors according to the updated finalization behavior model (\S\ref{sec_behavior_model}), or if new bugs are detected, the original transaction before disguising will be stored in the transaction seed pool to enrich the set of interesting transaction seeds (Lines 14-15).

\noindent
\textbf{Transaction mutation.}
Inspired by existing studies~\cite{yang2021finding,perez2019broken}, we enable the mutation of the execution context of transactions by utilizing contract creation transactions~\cite{solidity}.
The execution context of transactions refers to the opcode sequences executed on EVM during the transaction execution.
Different from other transactions, callee addresses of contract creation transactions are specified as \texttt{null}.
While transaction execution, contract creation transactions will execute the opcode sequences provided by transaction senders in the input data field of the transaction metadata~\cite{wood2014ethereum, solidity}.
Therefore, by mutating opcode sequences and specifying them in the input data field of contract creation transactions, we can alter the execution context of the corresponding contract creation transactions.

\begin{algorithm}[t!]
\caption{Behavior guided transaction fuzzing process.}
\label{alg_fuzzing}
\small{
\KwIn{\emph{zkRollup}: Polygon zkRollup testnet}
\KwIn{\emph{Seeds}: Initial transaction seeds}
\KwOut{\emph{Bugs}: Finalization failure bugs}

\emph{Model}, \emph{Bugs} $\leftarrow$ \{\} \\
\emph{zkRollup}.\texttt{Setup}() \\

\While{True}{
    \emph{tx} = \texttt{RandomSelect}(\emph{Seeds}) \\
    \emph{tx$_{m}$} = \texttt{Mutation}(\emph{tx}) \\
    \emph{tx$_{inj}$} = \texttt{LogicInjection}(\emph{tx$_{m}$}) \\
    \If{\texttt{Random}(0, 1) $>$ $P$}{
        \emph{feedback} = \emph{zkRollup}.\texttt{SendToRPC}(\emph{tx$_{inj}$}) \\
    } \Else{
        \emph{feedback} = \emph{zkRollup}.\texttt{SendToForcedBatch}(\emph{tx$_{inj}$}) \\
    }
    \emph{Model}.\texttt{UpdateModel}(\emph{feedback}) \\
    \emph{newBugs} = \texttt{BugOracle}(\emph{zkRollup}) \\
    \emph{Bugs}.\texttt{append}(\emph{newBugs}) \\
    \If{\texttt{IsTxInteresting}($Model$) or ($newBugs$ $\neq$ $\emptyset$)}{
        \emph{Seeds}.\texttt{append}(\emph{tx$_{m}$})
    }
}
}
\end{algorithm}

Our mutation strategies can be divided into two types, i.e., basic block mutation and opcode mutation, according to the granularity of the mutation strategies on the opcode sequences~\cite{zhou2018erays}.

\noindent
$\bullet$
\textit{Basic block mutation.} We prepare a corpus of basic blocks extracted from the bytecode of contracts deployed on Polygon zkRollup mainnet.
To mutate an opcode sequence, we first parse it into a list of basic blocks~\cite{zhou2018erays}. We then employ three mutation strategies at the granularity of basic blocks, i.e., insertion, deletion, and replacement, to mutate the list of basic blocks.
For insertion, we add a basic block from either the corpus or the list itself into the list.
For deletion, we remove a basic block from the list.
For replacement, we replace a basic block in the list with another basic block from either the corpus or the list itself. 
After mutating the list of basic blocks, we update the corpus with the basic blocks from the mutated list.

\noindent
$\bullet$
\textit{Opcode mutation.} We prepare a corpus of opcodes according to the specification of EVM opcodes~\cite{wood2014ethereum}.
To mutate an opcode sequence, we employ three mutation strategies at the granularity of opcodes, i.e., insertion, deletion, and replacement, to mutate the opcode sequence.
For insertion, we add an opcode from the corpus into the opcode sequence.
For deletion, we remove an opcode from the opcode sequence.
For replacement, we replace an opcode in the opcode sequence with another opcode from the corpus.
Additionally, since \texttt{PUSH} opcodes (\texttt{PUSH1-PUSH32}) require sequences of bytes ranging from 1 to 32 bytes in the opcode sequence as operands to push onto the stack, we also prepare a corpus of operands for \texttt{PUSH} opcodes which are extracted from the bytecode of contracts deployed on Polygon zkRollup mainnet.
When inserting, deleting, or replacing a \texttt{PUSH} opcode, we correspondingly adjust the associated operand from our corpus of operands.

\begin{figure}[t]
	\centering
	\includegraphics[width=0.97\linewidth]{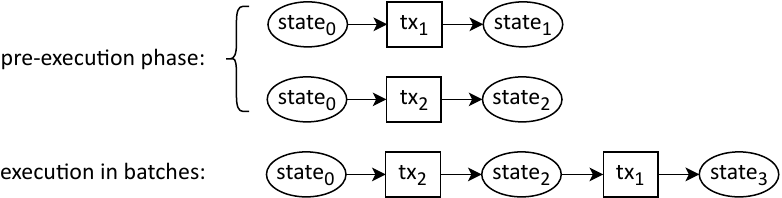}
	\caption{Between the pre-execution phase and execution in batches, $tx_1$ is executed on different blockchain states, thereby resulting in different state transitions.}
	\label{fig_logic_injection}
\end{figure}

\subsection{Logic injection}
\label{sec_logic_injection}
As mentioned in~\S\ref{sec_back_archi}, Txpool of Polygon zkRollup employs a pre-execution phase to filter out as many transactions as
possible that can trigger errors in any component before they are processed.
During the pre-execution phase, Txpool forks the current state of L2 blockchain, and executes transactions on the forked blockchain state to determine if they trigger any errors.
Our logic injection technique is designed to disguise the mutated transactions in order to bypass the pre-execution checks of Txpool, thereby ensuring that these transactions are processed by the finalization process.

A key observation inspiring the logic injection technique is that the blockchain state used to execute transactions during the pre-execution phase does not always match the state when these transactions are finally processed after being assembled into batches.
We use Fig.~\ref{fig_logic_injection} to illustrate this observation.
Assume there are two transactions, $tx_1$ and $tx_2$, to be included in Txpool, and the current blockchain state is denoted as $state_0$.
During the pre-execution phase, both $tx_1$ and $tx_2$ are executed on the forked blockchain state $state_0$, leading to state transitions to $state_1$ and $state_2$, respectively. 
After the pre-execution phase, $tx_2$ and $tx_1$ are included in the same batch in this sequence and executed starting from $state_0$.
Consequently, $tx_1$ executes on the blockchain state altered by $tx_2$, leading to a different state transition to $state_3$, which differs from its state transition ($state_1$) in the pre-execution phase.

By leveraging this observation, the logic injection employs three steps to successfully disguise a mutated transaction $tx_{m}$ for bypassing the pre-execution checks of Txpool.

\noindent
\textbf{-- Step 1:} We prepare a transaction $tx_c$ to deploy a contract $SC$ on L2 blockchain. The contract $SC$ contains a fallback function that only executes \texttt{revert()} operations. We assume that the transaction sender of $tx_c$ is $sender_1$ and the transaction nonce is $nonce_1$~\cite{wood2014ethereum}.
Hence, the contract address of $SC$ equals the first twenty bytes of the hash value of the RLP encoding of the sender's address and the nonce~\cite{wood2014ethereum}, i.e., $addr_{SC} = hash256(RLP(sender_1, nonce_1))$\emph{[0:20]}~\cite{wood2014ethereum}.
We would like to note that, before deploying $SC$, the return result for invoking $addr_{SC}$ is 1 (i.e., the default return value when callee does not exist). 
After deploying $SC$, the return result for invoking $addr_{SC}$ is 0 (i.e., the return value after executing \texttt{revert()}).

\noindent
\textbf{-- Step 2:} For the mutated transaction $tx_{m}$ where the corresponding opcode sequence to be executed is $op_{m}$, we inject control flow hijacking code into $op_{m}$ to alter its control flows. 
The opcode sequence after embedding the control flow hijacking code ($op_{hijack}$) is shown in Fig.~\ref{fig_step2_inject}.
Specifically, the opcode sequence in $op_{hijack}$ firstly invokes the contract $addr_{SC}$, and then checks the return result to determine the execution path after the \texttt{JUMPI} opcode in Line 2.
If the contract $addr_{SC}$ does not exist, the return result is 1, leading to the execution path to the \texttt{JUMPDEST} at Line 6.
Otherwise, if the contract $addr_{SC}$ exists, the return result is 0, and the opcode sequence after mutation will be executed as usual.

\noindent
\textbf{-- Step 3:} We generate a new contract creation transaction $tx_{inj}$ by specifying the opcode sequence $op_{hijack}$ in the input data field of $tx_{inj}$.
The transaction $tx_{inj}$ is signed by the sender $sender_1$ with the nonce incremented to $nonce_1 + 1$.
According to Polygon zkRollup documentation~\cite{polygon_zkevm_doc}, the transactions signed from the same account are processed in ascending nonce order.
Therefore, when included in the Txpool, both $tx_c$ and $tx_{inj}$ bypass the pre-execution checks, because the mutated opcode sequence $op_{m}$ does not execute.
Upon being included in the same batch, we can preserve that $tx_{inj}$ will execute after $tx_c$. Since $tx_{c}$ has deployed the contract $SC$, the return result for invoking $addr_{SC}$ will be 0.
This result enables $tx_{inj}$ to execute the mutated opcode sequence to explore potential bugs.
Besides, the logic injection successfully disguises the mutated transactions, enabling them to bypass the pre-execution checks of Txpool for subsequent processing in the finalization process.

\begin{figure}[t]
	\centering
	\begin{lstlisting}[language=Solidity,mathescape=true]
CALL // Invoke contract $addr_{SC}$
JUMPI // Jump to Tag #1 if the return result is 1
...
$\color{red}{\text{Original opcode sequences after mutation}}$ // $op_{m}$
...
JUMPDEST // Tag #1
RETURN
    \end{lstlisting}
	\caption{Opcode sequence with control flow hijacking code.}
	\label{fig_step2_inject}
\end{figure}

However, the logic injection does not always disguise the mutated transactions.
It fails if the transaction $tx_{c}$ has already been included in a batch to alter the blockchain state before the pre-execution checks for $tx_{inj}$.
In such cases, blockchain states for executing $tx_{inj}$ between the pre-execution phase and the execution in batches remain consistent regarding the states associated with $SC$.
However, the transaction $tx_{inj}$ may not be discarded by Txpool if it does not meet the specific conditions checked during the pre-execution phase.
If $tx_{inj}$ is discarded by Txpool, we regenerate a new pair of transactions, referred to as $tx_{c}^{\prime}$ and $tx_{inj}^{\prime}$, to ensure that $tx_{inj}^{\prime}$ can bypass the pre-execution checks for processing the mutated execution context in the subsequent finalization process.
We continue to generate new pairs of transactions until the newly generated $tx_{inj}$ is not discarded by Txpool during the pre-execution phase, or until a pre-specified threshold (e.g., 5) is reached.

\subsection{Bug analysis}
\label{sec_bug_analysis}

We develop bug oracles to monitor the testnet in real time to identify exceptional behaviors of components, thereby confirming the triggering of fast and hard finalization failure bugs.
As mentioned in~\S\ref{sec_system_model}, fast (resp. fard) finalization failure bugs violate the liveness of the fast (resp. hard) finalization process.
To precisely identify these bugs, we formally define the liveness properties of the fast and hard finalization processes as follows.

\noindent
\textbf{Definition 3}
(Liveness of fast finalization process).
There always exists a transaction $\exists tx_i \in TX$ such that eventually $\lozenge (tx_i \in B^j_2)$.
This liveness property preserves that there is always a transaction in Txpool that will eventually reach the fast finalized state (i.e., being processed in the L2 blockchain and included in a batch).

\noindent
\textbf{Definition 4}
(Liveness of hard finalization process).
There always exists a batch $\exists B^i_1 \in \mathcal{B}_{1}, (i > j)$ such that eventually $\lozenge p^i$.
This liveness property preserves that, after transactions in batch $B^j_1$ reach the hard finalized state, there is always a transaction in the fast finalized state will eventually reach the hard finalized state (i.e., the zk proof for the batch containing the transaction is generated).

According to the above definitions of liveness for the fast and hard finalization process, violations of their liveness can be categorized into three scenarios, i.e., i) a component within the corresponding finalization process crashes, ii) the execution of a component halts, and iii) a component enters in an unproductive state, e.g., the \sequencer{} continuously processing transactions without generating any batches.
Through monitoring program processes, and analyzing Polygon zkRollup execution logs and panic records, the first two scenarios can be straightforwardly detected~\cite{ma2023loki}.
However, confirming whether a component has entered an unproductive state presents challenges, as it requires analyzing component behaviors to identify sustained non-progress by any component.
Inspired by studies~\cite{chen2023tyr} that employ a decision time for determining liveness bugs in Ethereum clients, we confirm the unproductive state of a component when the component makes no progress in the finalization process for a period of decision time. The criteria for confirming the unproductive state for each component are as follows:

\noindent
-- RPC node is considered to have entered an unproductive state if it receives transactions continuously submitted from users, but fails to forward any transaction to Txpool.

\noindent --
Txpool is considered to have entered an unproductive state if it receives transactions continuously from RPC node, but fails to include any transaction in Txpool.

\noindent --
\sequencer{} is considered to have entered an unproductive state if there are transactions in Txpool, but \sequencer{} fails to produce any batch including these transactions.

\noindent --
\executor{} is considered to have entered an unproductive state if i) it continuously receives transactions forwarding from \sequencer{}, but fails to return execution results for any transaction, and ii) it continuously receives batches forwarding from \aggregator{}, but fails to generate witness for any batch.
Besides, the cryptographic proving backend within \executor{} is considered to have entered an unproductive state if it continuously receives witnesses forwarding from \executor{}, but fails to generate any zk proof for these batches.

\noindent -- \aggregator{} is considered to have entered an unproductive state if there are batches on L1 blockchain for which zk proofs are not generated, but \aggregator{} fails to forward any batch to \executor{}.

Since a component's unproductive state can be mistakenly determined, we employ a two-step process to confirm its unproductive state: i) We resign transactions that were successfully processed by the corresponding component, and prepare batches for these resigned transactions. ii) We then forward the corresponding transactions or batches to the component identified as potentially unproductive to confirm its current state.

\noindent
\textbf{Bug reproduction.}
To enhance the localization of detected finalization failure bugs and derive bug root causes, we develop a bug reproduction module. This module utilizes execution information and historical test transactions collected in the finalization behavior model to reproduce the finalization failure bugs.
By replaying historical test transactions, we can reproduce the finalization failure bugs for bug analysis. Additionally, by utilizing the collected execution information, we can trace the executed branches during the bug triggering to derive the root causes of identified bugs.
Furthermore, for the detected bugs that cause the components within the finalization process to crash, we utilize tools such as Sanitizers~\cite{sanitizers} to capture crash dump information. Subsequently, we analyze the causes of the crash using debugging tools with the collected crash dump information~\cite{ma2023loki}.
In addition to the identified bugs that cause the components to halt, thanks to the execution logs and panic records instrumented by Polygon zkRollup, we can analyze the root causes of these bugs by tracing these records~\cite{ma2023loki}.
\section{Implementation}
\label{sec_imple}

To evaluate the design of \tool{} and uncover finalization failure bugs of zero-knowledge layer 2 protocols, we implemented a prototype of \tool{} on Polygon zkRollup, encompassing all its released versions from the initiation of its mainnet on Mar. 27, 2023, to Feb. 24, 2024.
Additionally, we fork the latest Ethereum mainnet to serve as the layer 1 blockchain of our Polygon zkRollup testnet.
In the following, we delve into implementation details of \tool{}.

\noindent
\textbf{Initial state of testnet.} We reuse the genesis information of Polygon zkRollup to initiate our testnet. Besides, we randomly generate and maintain a list of accounts for signing transactions.
At the outset of testing, these accounts are endowed with initial funds (e.g., 5 ETH). Besides, incremental funds (e.g., 1 ETH) are allocated to them periodically during subsequent testing to ensure that they have sufficient funds to cover transaction fees, thereby preventing the signed transactions are preemptively discarded by Txpool.

\noindent
\textbf{Construction for initial transaction seeds.}
The initial transaction seeds are constructed by reusing the bytecode from randomly selected contracts deployed on Polygon zkRollup mainnet.
Specifically, we extract the contract bytecode and specify it as the data field in corresponding contract creation transactions.

\noindent
\textbf{Modifications in \rom{}.} As a virtual machine functionally equivalent to EVM, \rom{} terminates transaction execution when the destination of jump opcodes (\zkcode{JUMP} and \zkcode{JUMPI}) is not \zkcode{JUMPDEST}, as only \zkcode{JUMPDEST} is recognized as a valid jump target.
Moreover, according to the specification of Polygon zkRollup, \rom{} also terminates transaction execution if the memory location accessed by memory read/write opcodes (e.g., \zkcode{MLOAD} and \zkcode{MSTORE}) exceeds the limit of \zkcode{0x20000}.
Please note that, in transaction mutation, the opcode sequences and the list of basic blocks are randomly mutated (\S\ref{sec_fuzzing}). 
Therefore, the transaction mutation increases the chance of terminating transaction execution prematurely due to the two restrictions.
To address this issue, we disable the destination checking for \zkcode{JUMPDEST} of jump opcodes.
Furthermore, for memory accessing, we apply a modulo \zkcode{0x20000} operation to the actual memory locations accessed, thereby focusing transaction execution within a finite memory region and avoiding inefficient fuzzing due to boundless state space in memory~\cite{bohme2020boosting}.

\noindent
\textbf{Decision time setup.}
Decision time is used to confirm whether components have entered an unproductive state (\S\ref{sec_bug_analysis}) to confirm the triggering of bugs. %
Inspired by the observation that prolonged decision time impacts the fuzzing iteration efficiency while effectively reducing false positives, we adopt the decision time used in Tyr~\cite{chen2023tyr}, i.e., 72 seconds as six times of L1 block confirmation time.

\noindent
\textbf{Tradeoffs on fuzzing efficiency.} Throughout the finalization process, the most time-consuming task is generating zk proofs using the cryptographic proving backend in \executor{}. This time-consuming task leads to inefficiency in fuzzing iterations.
For example, the average transaction execution time is 0.56 milliseconds~\cite{chen2021forerunner}, whereas the zk proof generation process for a batch typically exceeds 2 minutes~\cite{zkevm_proof_time_cost} (at least 214,285 times the duration of a transaction execution). 
To enhance the efficiency in fuzzing iterations, we assume it is sufficient to ascertain the feasibility of generating a zk proof rather than actually generating the zk proof.
Under this assumption, once a witness is generated in \executor{}, we assign a random number to serve as the zk proof and submit this to the L1 blockchain.
This assumption is considered reasonable, because the actual process of generating a zk proof from a witness is conducted within well-established cryptographic tools~\cite{ben2018fast,ben2019scalable,belles2022circom,polygon_zkevm_doc}. However, this approach has limitations as it prevents the identification of bugs within the cryptographic tools that impede the finalization process.

\section{Evaluation}
\label{sec_evaluation}

\noindent We evaluate the bug-finding performance of \tool{} on Polygon zkRollup by answering the following four research questions. 
\textbf{RQ1:} \textit{How effective is \tool{} in identifying bugs?}
\textbf{RQ2:} \textit{Can \tool{} outperform baselines in terms of both bug finding capability and coverage?}
\textbf{RQ3:} \textit{How does \tool{}'s bug finding performance benefit from components like finalization behavior model and logic injection?}
\textbf{RQ4:} \textit{What insights can we derive from bug root causes?}
\textbf{RQ5:} \textit{Can \tool{} be extended to other L2 protocols?}

All our experiments are conducted on a 64-bit machine equipped with 64 CPU cores and 128 GB memory. 
We operate each Polygon zkRollup's component within a separate Docker container to mitigate resource contention. These containers are configured using the information provided in the Dockerfile from Polygon zkRollup's official repository~\cite{polygon_zkevm_node}.
We would like to note that \tool{} is the first tool designed to detect finalization failure bugs in zero-knowledge layer 2 protocols. Among existing blockchain fuzzing tools~\cite{yang2021finding,chen2023tyr,ma2023loki,fu2019evmfuzzer,kim2023etherdiffer,bano2022twins,wang2023understanding,beaconfuzz,evmlab}, there are no qualitative baseline tools to measure common fuzzing metrics like coverage.
To evaluate the effectiveness of \tool{}, we establish two baseline tools for comparison.
The first baseline tool, Fluffy-F, originates from a state-of-the-art blockchain bug-finding tool, Fluffy~\cite{yang2021finding}, to assess the extent to which \tool{} outperforms existing tools.
The second baseline tool, Fuzzer-Cover, exclusively employs the transaction fuzzer (\S\ref{sec_fuzzing}) of \tool{} guided by coverage information, excluding other core modules of \tool{} like the finalization behavior model (\S\ref{sec_behavior_model}) and logic injection technique (\S\ref{sec_logic_injection}).
Furthermore,
we embed bug oracles from \tool{} (\S\ref{sec_bug_analysis}) for the two baseline tools to enable their capability to detect finalization failure bugs.

\subsection{RQ1: Bug-finding capability}
\label{sec_rq1}

\begin{table*}[t]
  \small
  \centering
  \caption{Descriptions for the twelve zero-day finalization failure bugs identified by \tool{}}
  \resizebox{0.99\linewidth}{!}{
  \begin{tabular}{@{}ccl@{}}
  \toprule
  \textbf{\#} & \textbf{Type} & \textbf{Bug Description} \\ \midrule
  1    &   \textbf{FFF}       &    Inconsistent counter limits between \sequencer{} and \executor{} raises an out-of-counter error, causing the shutdown of \sequencer{}.             \\ 
  2    &   \textbf{FFF}       &        Transactions whose counters exceed the limits for batch assembly flood txpool, leading \sequencer{} to continuously reprocess them.        \\
  3    &   \textbf{FFF}       &        Transactions trigger the inconsistent calculations between \sequencer{} and \executor{} on transaction senders flood txpool.        \\
  4    &   \textbf{HFF}       &  Constraints for long division are unsatisfied when the data copying triggers an overflow, leading to the halt of proof generation.  \\ %
  5    &   \textbf{FFF}      &        \sequencer{} mishandles the error \zkcode{ZKR\_SM\_MAIN\_ADDRESS} raised by \executor{}, causing txpool to be flooded.         \\
  6    &   \textbf{HFF}       &  \executor{} raises an unexpected error due to incorrectly tallying the memory align operations, causing the halt of proof generation.             \\
  7    &   \textbf{HFF}      &  Proof generation halts due to a memory leak caused by the neglect in deallocating the initialized finite field variables.           \\
  8    &   \textbf{HFF}      &  \executor{} incorrectly marks the last 2048 bytes of memory region as inaccessible, causing errors that halt future proof generation.      \\ 
  9    &  \textbf{FFF}  &   \sequencer{} halts producing batches and L2 blocks when receiving a transaction with metadata containing an odd-length hex string.      \\
  10    &  \textbf{FFF}   &   \sequencer{} fails to handle inaccessible issues of state database, causing a mutex to be double unlocked and leading to its crash.      \\
  11    &   \textbf{HFF}    &  Flawed memory boundary check in \executor{} causes unexpected errors, thereby preventing proof generation for all future batches.        \\ %
  12    &   \textbf{HFF}      &  The case of integer division is overlooked in long division, halting proof generation due to unsatisfied constraints for a remainder 0.  \\
  \bottomrule
  \end{tabular}
  }
  \label{tab_bugs}
  \vspace{6pt}
\end{table*}

We conduct experiments by running \tool{} on each released version of Polygon zkRollup for 24 hours to find finalization failure bugs. In total, \tool{} found twelve finalization failure bugs that are previously unknown.
Among these, six bugs are classified as fast finalization bugs, and six bugs are categorized as hard finalization bugs, each affecting the finalization process differently.

We summarize the descriptions of the twelve finalization failure bugs in Table~\ref{tab_bugs}.
At the time of writing, all the twelve finalization failure bugs have been confirmed by official Polygon zkRollup teams, evidencing the validity of the twelve bugs detected by \tool{}.

\noindent
\textbf{Case study.} 
We illustrate how hard finalization failure bugs disrupt the hard finalization process by utilizing the Bug \#11 in Table~\ref{tab_bugs}, and detail other our identified bugs in Appendix B.
This approach aims to provide a comprehensive understanding of the fast and hard finalization failure bugs and their security impacts.

\begin{figure}[t]
	\centering
	\includegraphics[width=0.65\linewidth]{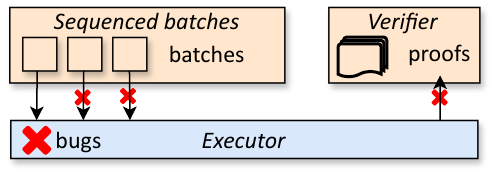}
	\caption{%
 Errors in \executor{} halt the proof generation, preventing the generation of proofs for all future batches.}
	\label{fig_example_figure}
\end{figure}

As shown in Fig.~\ref{fig_example_code}, \executor{} raises an error \zkcode{ZKR\_SM\_MAIN\_ADDRESS} when the accessed maximum memory offset exceeds the checked boundary (\zkcode{0x10000}), whereas the actual valid memory boundary should be \zkcode{0x20000}~\cite{polygon_zkevm_doc}. 
Consequently, in the case of a valid transaction where the accessed maximum memory offset surpasses the checked boundary (\zkcode{0x10000}), \executor{} halts its execution due to the encountered error.
When such a transaction is processed within \sequencer{}, it will be directly discarded due to the encountered errors within \executor{}.
However, if a transaction triggering this bug is submitted to Forced batches, \executor{} will be forced to proceed with its execution and zk proof generation after this transaction is included in sequenced batches on L1 blockchain.
In this scenario, due to the raised errors, \executor{} halts the proof generation process for this transaction.
Consequently, as shown in Fig.~\ref{fig_example_figure}, the batch containing this transaction fails to obtain its zk proof, leading to the prevention of generating valid proofs for all subsequent batches.

\begin{figure}[t]
	\centering
	\begin{lstlisting}[language=C++]
if (addrRel>=0x20000 || ((rom.line[zkPC].isMem==1) && (addrRel >= 0x10000))){ /*addrRel >= 0x20000*/
  cerr << "Error: addrRel too big addrRel=" << addrRel << " step=" << step << " zkPC=" << zk
  proverRequest.result = ZKR_SM_MAIN_ADDRESS;
  return;}
    \end{lstlisting}
	\caption{Code snippets from a hard finalization failure bug. 
    }
	\label{fig_example_code}
\end{figure}

\noindent
\textbf{Summary.} 
FFF bugs prevent the production of batches and L2 blocks, causing L2 transactions cannot reach fast-finalized states. HFF bugs inhibit the generation of proofs for subsequent batches, causing L2 transactions cannot reach hard-finalized states.

\subsection{RQ2: Can \tool{} outperform baselines?}
\label{sec_rq2}

In this section,
we run the two baselines, i.e., Fluffy-F and Fuzzer-Cover, on the same experimental environments as \tool{} in~\S\ref{sec_rq1}, to examine whether \tool{} can outperform the two baselines in terms of finding bug capability and coverage.

\noindent
\textbf{Comparison on finding bugs.}
In our experiments of Fluffy-F and Fuzzer-Cover, the two baselines cannot identify any new finalization failure bugs. Out of the twelve bugs identified by \tool{}, the baselines can only successfully detect three of them, i.e., Bug\#3, Bug\#9, and Bug\#10, as shown in Table~\ref{table_bug_find_com}.
To facilitate future research on detecting finalization failure bugs, we conduct an investigation into why the two baselines successfully detect the three bugs while failing to trigger the remaining nine bugs.

\noindent
$\bullet$
\textit{Reasons for successfully detecting bugs.}
The three bugs detected by baselines originate from \sequencer{}'s logic bugs while handling incoming transactions at the beginning of the finalization process.
Triggering the three bugs does not require satisfying deep path conditions or involving complex interactions between L2 components. For example, to trigger Bug\#10, baselines only need to generate a test transaction with metadata containing an odd-length hex string.

\noindent
$\bullet$
\textit{Reasons for failing to detect bugs.} The reasons are threefold.

\noindent
i) Both baselines cannot be aware of inherent behaviors and states of L2 components within the finalization process. Consequently, they rely solely on the random mutation of transactions guided by coverage information as feedback. This limitation prevents them from generating test transactions capable of triggering diverse behaviors of L2 components to expose bugs. For example, Bug\#1 necessitates the generation of transactions incurring various Counters.
However, the feedback mechanisms of both baselines do not capture the internal behaviors of \executor{} during transaction execution.

\noindent
ii) Transactions generated by the two baselines 
can be preemptively discarded by Txpool when the transactions access invalid memory regions during pre-execution phases. This is because the baselines do not incorporate the modifications of \tool{} in \rom{} for relocating accessed memory locations (\S\ref{sec_imple}). This limitation causes \executor{} 
prematurely terminate the transaction execution due to invalid operations, causing the transactions to be discarded by Txpool and preventing the transactions from triggering bugs.

\noindent
iii) The premature termination of \executor{} when executing transactions generated by the two baselines cannot trigger exceptional errors within \executor{}. Therefore, even if these transactions are submitted to Forced batches on L1 blockchain, the generation of zk proofs for batches containing these transactions cannot halt the proof generation process to trigger hard finalization failure bugs.

\noindent
\textbf{Comparison on coverage.}
To further assess the efficacy of \tool{} and the baselines in terms of bug detection capability, we analyze the trends of their coverage growth over time. 
Since we conduct multiple rounds of experiments on distinct versions of Polygon zkRollup, we average the coverage information of each tool over time.
Overall, after 24 hours of experimentation, Fuzzer-Cover and Fluffy-F cover 13,089 and 14,219 branches on average.
In comparison, \tool{} covers 17,182 branches on average, surpassing Fluffy-F by 20.8\% and Fuzzer-Cover by 31.3\% in branch coverage, respectively.
The main reason that \tool{} outperforms baselines in coverage can be attributed to the utilization of the finalization behavior model.
The model enables \tool{} to continuously select and mutate test inputs to trigger diverse behaviors in Polygon zkRollup components, thus enhancing coverage performance.
We refer readers to Appendix C for details on the trends of coverage growth over time for each tool.

\begin{table}[t]
\centering
\caption{Finalization failure bugs identified by two baselines}
\resizebox{0.56\linewidth}{!}{
\begin{tabular}{ll}
\toprule
\textbf{Baselines} & \textbf{Identified bugs} \\ \midrule
Fluffy-F &   Bug\#3, Bug\#9, Bug\#10            \\
Fuzzer-Cover  &      Bug\#3, Bug\#9, Bug\#10           \\ \bottomrule
\end{tabular}
}
\label{table_bug_find_com}
\end{table}

\begin{table}[b]
\centering
\caption{Finalization failure bugs identified by two variants}
\resizebox{0.85\linewidth}{!}{
\begin{tabular}{ll}
\toprule
\textbf{Variants} & \textbf{Identified bugs} \\ \midrule
\tool{}-OmitM &  \begin{tabular}[c]{@{}l@{}}Bug\#2, Bug\#3, Bug\#5, Bug\#9, Bug\#10\end{tabular}    \\ \midrule
\tool{}-OmitI  &  \begin{tabular}[c]{@{}l@{}}Bug\#1, Bug\#3, Bug\#4, Bug\#6, Bug\#7,\\ Bug\#8, Bug\#9, Bug\#10, Bug\#11, Bug\#12\end{tabular}  \\ \bottomrule
\end{tabular}
}
\label{table_variant_com}
\end{table}

\noindent
\textbf{Consideration for selecting baselines.}
When determining baseline tools, we initially selected several existing tools, such as Fluffy. We then conducted a preliminary evaluation to compare these potential baselines with the baseline built by us, Fuzzer-Cover, the naive coverage-based fuzzing tool, in terms of coverage to further filter the baselines for evaluation. After this step, most of the initially selected baselines, including Tyr and LOKI, were filtered out. This is because their methodology and system model are unsuitable for detecting bugs in Polygon zkRollup. For example, Tyr and LOKI detect bugs in blockchain consensus protocols. However, Polygon zkRollup operates with components without a consensus mechanism. Additionally, since these tools focus on mutating P2P messages, e.g., \zkcode{getBlockHeader} requests, they cannot efficiently mutate transaction execution contexts to explore different paths. %

\subsection{RQ3: Effects of components in \tool{}}

We develop two inferior variants of \tool{}, aimed at evaluating how the bug-finding performance of \tool{} benefits from the finalization behavior model and the logic injection technique.
Our experimental environments for running the two variants of \tool{} are the same as \tool{} in~\S\ref{sec_rq1}.

\noindent
\textbf{Variant 1.} \tool{}-OmitM excludes the finalization behavior model (\S\ref{sec_behavior_model}) from \tool{}, and utilizes the coverage information to guide the transaction fuzzer like Fuzzer-Cover in~\S\ref{sec_rq2}.

\noindent
\textbf{Variant 2.} \tool{}-OmitI excludes the logic injection (\S\ref{sec_logic_injection}) from \tool{}, and directly submits mutated transactions to testnet.

According to our results displayed in Table~\ref{table_variant_com}, out of the twelve bugs identified by \tool{}, \tool{}-OmitI successfully detects ten of them, while \tool{}-OmitM can detect five of them.

Without the finalization behavior model to guide fuzzing, \tool{}-OmitM relies solely on mutating transactions guided by coverage information as feedback, similar to Fuzzer-Cover. Hence, both \tool{}-OmitM and Fuzzer-Cover successfully detect the three bugs, i.e., Bug\#3, Bug\#9, and Bug\#10, that do not require satisfying deep path conditions or involving complex interactions between L2 components to trigger.
Moreover, with logic injection, \tool{}-OmitM can additionally detect two bugs, Bug\#2 and Bug\#5, compared to Fuzzer-Cover, because the corresponding test transactions generated by \tool{}-OmitM will not be preemptively discarded by the Txpool during the pre-execution checks.

\begin{table}[t]
\centering
\caption{Bug detection accuracy on different decision time}
\resizebox{0.93\linewidth}{!}{
\begin{tabular}{lcccccccc}
\toprule
\textbf{Decision time (s)} & 1 & 3 & 6 & 12 & 24 & 48 & 72 & 120 \\ \midrule
\textbf{False positives} & 30 & 14 & 3 & 0 & 0 & 0 & 0 & 0 \\ 
\textbf{False negatives} & 0 & 0 & 0 & 0 & 0 & 0 & 0 & 0 \\
\textbf{True positives} & 12 & 12 & 12 & 12 & 12 & 12 & 12 & 12  \\ \bottomrule
\end{tabular}
}
\label{table_decision_time}
\end{table}

Without the logic injection technique, \tool{}-OmitI fails to trigger the two fast finalization failure bugs, Bug\#2 and Bug\#5, because the corresponding test transactions generated by \tool{}-OmitI are preemptively discarded by the Txpool during the pre-execution checks.
Despite the two undetected bugs, under the guidance of the finalization behavior model, \tool{}-OmitI successfully triggers and detects the remaining ten bugs.

\noindent
\textbf{Impact of decision time on bug detection accuracy.}
As mentioned in~\S\ref{sec_bug_analysis}, we employ a decision time to confirm the unproductive state of components for identifying finalization failure bugs.
To examine the impact of decision time on the accuracy of \tool{}'s bug finding, we run \tool{} in the same experimental environments as~\S\ref{sec_rq1} by incorporating various values of decision time, and collect the bugs identified by it for the accuracy analysis.

According to our results displayed in Table~\ref{table_decision_time}, extending decision time decreases the incidence of false positives. When decision time is minimal (e.g., 1 second), components are more likely to be mistakenly confirmed as unproductive if they are engaged in processing transactions that require extended execution times.
Moreover, prolonged decision time does not influence the precision of \tool{} in identifying
true bugs (i.e., false negatives and true positives). 
This is because once finalization failure bugs are triggered, these bugs will cause the affected components to remain in a persistently unresponsive state or even lead to an execution crash.
However, when solely employing our bug oracles as an online incident detection tool within the Polygon zkRollup, a prolonged decision time delays the developers' awareness and response to the triggered bugs, potentially impacting the timeliness of interventions.

\noindent
\textbf{Summary.} Finalization behavior model is critical for detecting both fast and hard finalization failure bugs, because it guides the fuzzer to explore varied behaviors of L2 components. The logic injection is critical for detecting fast finalization failure bugs by enabling test transactions to bypass pre-execution checks.
Prolonged decision time reduces the incidence of mistakenly identifying non-bugs.

\subsection{RQ4: Lessons from bug root causes}
\label{sec_root_cause}

Following an in-depth investigation of bug root causes, i.e., (i) categorizing all bugs according to their root causes, and (ii) examining each type of root cause to summarize corresponding insights, we derive the following findings. 
We believe our insights will not only enhance the security of Polygon zkRollup, but also provide valuable knowledge for developers on other L2 protocols.

\noindent
$\bullet$
\textit{\textbf{From Bug\#1-3,5: Inconsistent behaviors among components lead to unexpected issues.}}
In Polygon zkRollup, tasks in the finalization process typically require the collaboration of multiple components.
However, due to the high coupling nature between these components, inconsistent checking and processing between them can result in unexpected issues.
For example,
the assembly of a batch requires coordination among \sequencer{}, \executor{}, and \rom{}.
Among them, \sequencer{} is responsible for bundling and ordering transactions into batches, \executor{} manages the execution of transactions and the generation of execution witnesses, while \rom{} interprets each operation within the transaction execution.
However, inconsistent checks between them will cause each component to produce different judgments about whether transactions and batches are valid, thereby leading to unexpected errors.

\noindent
$\bullet$
\textit{\textbf{From Bug\#1,2,5: Transaction execution environments differ between the inclusion in txpool and batches.}}
Transactions that trigger execution errors during the pre-execution phase will be discarded by the transaction pool.
However, the transaction execution environment during pre-execution may differ from that during batch assembly.
Hence, filtering out all error-triggering transactions is challenging, potentially leading to execution errors in subsequent processing within Polygon zkRollup. 
It is worth noting that our logic injection technique aims to generate transactions capable of bypassing pre-execution checks more efficiently, thereby enhancing our tool's effectiveness in triggering bugs. Nonetheless, when exploited by malicious attackers, this technique can also empower attackers to craft attack transactions that more reliably evade pre-execution checks, posing a threat to Polygon zkRollup's security.

\noindent
$\bullet$
\textit{\textbf{From Bug\#4,6-8,11,12: Forced batches: a double-edged sword}.}
The primary purpose of introducing Forced batches is to maintain the liveness of L2 blockchain. 
In instances where L2 components become unresponsive or behave maliciously, users can still ensure their L2 transactions are sequenced onto L1 blockchain through Forced batches.
Subsequently, users can obtain zk proofs for their transactions, ensuring their transactions reach a hard-finalized state.
Taking cross-chain transfers as an example, a transaction that reaches a hard-finalized state via Forced batches will be recognized as valid by the cross-chain bridge, allowing subsequent transfers to be processed accordingly.
However, the implementation of Forced batches also introduces new vulnerability surfaces within Polygon zkRollup. 
For example, there are some transactions that may trigger execution errors in \executor{}.
Normally, these transactions, when processed by \sequencer{}, would prematurely trigger corresponding execution issues, preventing them from being submitted to L1 blockchain.
However, by interacting with Forced batches, these transactions will force \executor{} to process them, thus triggering errors to impede the generation of valid zk proofs. 
Consequently, such cases will prevent the generation of zk proofs for all subsequent batches, thereby disrupting the hard finalization process entirely.

\noindent
$\bullet$
\textit{\textbf{From Bug\#2,3,5: Mitigating the impact of bugs through runtime protection on txpool.}}
Preemptively eliminating all potential bugs during the development phase is impractical. 
However, given that a significant number of bugs obstruct the fast finalization process by flooding transaction pool,
developers could employ runtime checks on transaction pool.
This approach would ensure the transaction pool's proper function or enable rapid recovery from any abnormal deviations, thereby mitigating the impact of such bugs.

\subsection{RQ5: Generality to other L2 protocols}
\label{sec_generality}

Other zero-knowledge and optimistic L2 protocols, such as Scroll zkRollup~\cite{scroll} and Optimism Rollup~\cite{Optimism}, employ the transaction finalization mechanism with the two stages that are similar to those used in Polygon zkRollup.
This enables the potential of \tool{} to find finalization failure bugs in these L2 protocols.
In this section, we explore the generality of our approach by evaluating whether \tool{} can identify finalization failure bugs in other L2 protocols.

Tailoring \tool{} to the unique design of other L2 protocols necessitates additional efforts.
Specifically, the procedure of extending \tool{} to other L2 protocols involves three steps:
First, \tool{} resolves unique components within the finalization process of other L2 protocols. For example, Optimism Rollup introduces \validator{} rather than \aggregator{} for conducting interactive fraud proofs for batches in the hard finalization process.
Second, \tool{} updates new behavior models tailored to corresponding L2 protocols to characterize how their unique components are engaged in the finalization process.
Third, \tool{} resolves new metrics for measuring transaction executions on other L2 protocols, such as Counters, for characterizing their transaction executions.

\noindent
\textbf{Generality to other zero-knowledge L2 protocols.}
We extend \tool{} to Scroll zkRollup, another prominent zero-knowledge L2 protocol, to explore the generality of our approach across other zero-knowledge L2 protocols.
For this evaluation, we reuse the experimental environments previously employed by \tool{} on Polygon zkRollup, running \tool{} on the latest versions of Scroll zkRollup to detect finalization failure bugs.
Through this process, we identify a previously unknown fast finalization failure bug in Scroll zkRollup (cf. Appendix D), thereby demonstrating the effectiveness of \tool{} in detecting finalization failure bugs on other zero-knowledge L2 protocols. At the time of writing, this bug has been confirmed and fixed by the official Scroll zkRollup team.

\noindent
\textbf{Generality to other optimistic L2 protocols.}
We further extend \tool{} to Optimism Rollup, a leading optimistic Layer 2 (L2) protocol, to explore the generality of our approach on other optimistic L2 protocols.
Similar to the evaluation on Scroll zkRollup, we run \tool{} on the latest versions of Optimism Rollup to detect finalization failure bugs.
However,
during this process, we do not find any unknown finalization failure bugs in Optimism Rollup. There are two possible reasons for this:
First, our preliminary extension of \tool{} on Optimism Rollup may have missed several unique interfaces and mechanisms, limiting its ability to trigger finalization failure bugs associated with these specific designs. Second, Optimism Rollup has undergone extensive development, testing, and auditing over several years. As a result, its transaction finalization process, as a core functionality, has been thoroughly examined, making it difficult to uncover bugs in the latest stable versions.

To further explore the potential of \tool{} in detecting finalization failure bugs on optimistic L2 protocols, we build a dataset of known finalization failure bugs on Optimism Rollup with the associated client versions, and examine the bug-finding capability of \tool{} on this dataset.
To ensure the validity of the selected bugs, we collect known bugs from the audit reports acknowledged in the official repository~\cite{Optimism-audit}. 
Please note that there is no specific category for finalization failure bugs in these audit reports.
Besides, auditors typically focus on exposing diverse security issues in the code repository.
Therefore, we choose to manually filter the finalization failure bugs from the reported issues in these audit reports. Specifically, we employ three authors to analyze the reported bugs, focusing on their root causes and impact, and discuss to confirm whether a reported bug is a finalization failure bug. Once a bug is confirmed, we investigate the pull requests fixing the bugs, as referenced in the audit reports, to track the client versions affected by the bugs. As a result, we successfully determine two fast finalization failure bugs from the audit reports (cf. Appendix D).

We conduct the evaluation to the buggy versions of Optimism Rollup
using the same experimental configurations as above. Through this process, we successfully reproduce and reconfirm the two bugs identified in our dataset, demonstrating the potential of \tool{} in detecting finalization failure bugs on optimistic L2 protocols.

\noindent
\textbf{Summary.} 
\tool{} can both uncover unknown finalization failure bugs and reconfirm known finalization failure bugs in other zero-knowledge and optimistic L2 protocols, 
\section{Discussion}
\label{sec_discuss}
\noindent
\textbf{Limitations on generality evaluation.}
In \S\ref{sec_generality}, we explore the generality of \tool{} to other L2 protocols, by examining its bug-finding capability in Scroll zkRollup and Optimism Rollup. Although \tool{} can uncover unknown and reconfirm known bugs in these protocols, our evaluation has limitations in two aspects. First, our constructed bug dataset on Optimism Rollup may lack completeness, potentially missing bugs reported in other audit reports or through other channels like blogs. Second, the L2 protocols we selected for evaluation are not exhaustive, as there are other L2 protocols~\cite{Taiko,Arbitrum}. %
Notably, \tool{} is designed to detect finalization failure bugs in Polygon zkRollup. Therefore, we choose to explore its potential to extend to other L2 protocols rather than completely detect all potential bugs in them. Additionally, extending \tool{} to other L2 protocols requires additional manual efforts. Hence, we selected two representative zero-knowledge and optimistic L2 protocols to demonstrate the generality of our approach to other L2 protocols.

\noindent
\textbf{Bug categories in Polygon zkRollup.}
We focus on finalization failure bugs that thoroughly disrupt the liveness of transaction finalization process in Polygon zkRollup, such as preventing it from confirming new transactions.
To provide a clearer picture of the scope of our study, we list four other types of bugs in Polygon zkRollup in the following, based on their impact:
(i) bugs that lead to double-spending~\cite{sguanci2021layer,sun2024doubleup}, (ii) soundness bugs (e.g., under-constrained bugs) that cause valid zk proofs to be generated for invalid state transitions~\cite{wen2023practical}, (iii) bugs that delay user transactions~\cite{medium_offchainlabs_rollup_delay}, 
and (iv) griefing attacks that cause the freezing and loss of user funds~\cite{Griefing}.
However, these bugs do not directly compromise the liveness of Polygon zkRollup, and are thus out of the scope of our study.

\noindent
\textbf{Selection for new behaviors.}
The finalization behavior model in~\S\ref{sec_behavior_model} employs behavior data of components within the finalization process to characterize the finalization behaviors.
These behavior data are selected based on how these components are engaged in the finalization process. 
This raises threats to our validity, as some behaviors are not as interesting as others in guiding the subsequent iterations. However, it is non-trivial to distinguish which behaviors are more interesting, because there are no qualitative metrics.

\noindent
\textbf{Responsible bug disclosure and mitigations.}
\label{sec_bug_disclose}
We reported all the bugs we revealed to the Polygon zkRollup team and the Scroll zkRollup team via the Immunefi platform~\cite{immunefi}.
At the time of writing, both the two official teams have confirmed all our bug reports.
In addition, the two official teams have awarded us corresponding bug bounties to acknowledge and reward our contributions in revealing these bugs.
While reporting the bugs, we also provided feasible mitigations, and the official developers followed our advice to fix all the bugs accordingly.
Furthermore, we also investigated the root causes of these bugs, as detailed in~\S\ref{sec_root_cause}.
During the discussion with the two official teams, we received positive feedback from the official developers on the revealed bugs, proposed mitigations, and our analysis of their root causes.

\noindent
\textbf{Ethical concerns.}
\label{sec_ethical}
When evaluating the effectiveness of \tool{}  in identifying finalization failure bugs, we confined the implementation of \tool{} to our local environment, ensuring that experiments did not involve external parties and the mainnet of Polygon zkRollup.
When reporting bugs, we exclusively disclosed the bug details to Polygon zkRollup and Scroll zkRollup teams through the Immunefi platform, and each team received reports only about the bugs relevant to them~\cite{immunefi}.
Before the bugs were fixed in the official repository, we did not disclose any bug details to the public.
Moreover, when suggesting bug mitigations, we informed Polygon zkRollup team that the mitigations might not completely resolve the bugs and could potentially introduce new attack surfaces.

\section{Related work}
\label{sec_relate_work}

\noindent
\textbf{Vulnerabilities in layer 2 protocols.}
Current research on layer 2 security focuses on uncovering attacks targeting Bitcoin lighting network and Ethereum optimistic rollup protocols. 
Malavolta et al.~\cite{malavolta2018anonymous} propose the wormhole attack against payment-channel networks, enabling dishonest users to steal the payment fees from honest participants along the payment path.
Herrera-Joancomart{\'\i} et al.~\cite{herrera2019difficulty} unveil an attack strategy to disclose a payment channel's balance in Bitcoin lightning network. Their core idea involves initiating multiple payments, ensuring that none of them is finalized to minimize the attack economic cost.
Riard et al.~\cite{riard2020time} introduce the time-dilation attack, which prolongs the time interval for victims to be aware of new blocks by network isolation and block delivery delays.
By leveraging the time-dilation attack, an adversary can steal funds from victims' payment channels.
Koegl et al.~\cite{koegl2023attacks} collect a list of known attacks (e.g., sybil attack) on Ethereum layer 2 protocols and illustrate their impact.
Offchain Labs~\cite{medium_offchainlabs_rollup_delay} discuss the security impact of delay attacks on Ethereum optimistic rollup protocols, and evaluate several mitigations against the delay attacks.

\noindent
\textbf{Blockchain bug detection.}
Researchers proposed various approaches to detect bugs within blockchain systems. These studies can be divided into three types based on the bug location, i.e., consensus protocols, blockchain infrastructure, and smart contracts.

\noindent
\textit{-- Consensus Protocols.} Consensus protocols play a crucial role in coordinating nodes to reach an agreement on the latest blockchain state. Bugs in consensus protocols will threaten the validity and consistency of blockchain systems~\cite{chen2023tyr,zou2023optimized}.
LOKI~\cite{ma2023loki} serves as a consensus protocol fuzzing framework to detect consensus memory-related and logic bugs.
Fluffy~\cite{yang2021finding} detects consensus failure bugs originating from Ethereum virtual machine by conducting multi-transaction differential fuzzing between Geth and OpenEthereum. 
Tyr~\cite{chen2023tyr} detecting consensus failure bugs in blockchain systems based on a behavior divergent model.
Twins~\cite{bano2022twins} is a unit test generator to perform Byzantine attack on Diem blockchain.

\noindent
\textit{-- Blockchain infrastructure.} Blockchain infrastructure consists of components that provide the fundamental functionalities of blockchain systems, such as Ethereum virtual machine, RPC services, and transaction pool. Bugs in blockchain infrastructure directly lead to unexpected execution errors.
EVMLab~\cite{evmlab} and EVMFuzzer~\cite{fu2019evmfuzzer} identify flaws in Ethereum virtual machine by comparing execution inconsistencies among multiple implementations of Ethereum virtual machine.
Beaconfuzz~\cite{beaconfuzz} employs libFuzzer~\cite{libfuzzer} to detect bugs across different Ethereum 2.0 implementations via coverage-guided differential fuzzing.
Phoenix~\cite{ma2023phoenix} designs context-sensitive chaos testing to detect node unrecoverable and data unrecoverable issues for blockchain clients.
EtherDiffer~\cite{kim2023etherdiffer} identifies implementation bugs, e.g., crash and denial-of-service bugs, for blockchain RPC services.
MPFuzz~\cite{wang2023understanding} finds asymmetric DoS bugs by symbolically
exploring Ethereum transaction pool's state space.

\noindent
\textit{-- Smart contract.} Smart contracts enable blockchain users to engage in various purposes. Bugs in contracts can lead to financial losses for blockchain users and may even impact the security of blockchain systems.
Researchers have proposed various program analysis methods to detect diverse contract bugs and vulnerabilities.
For example, Mythril~\cite{Mythril} and Oyente~\cite{luu2016making} use symbolic execution to explore execution flows for detecting vulnerabilities like reentrancy.
Smartian~\cite{choi2021smartian} and Echidna~\cite{grieco2020echidna} utilize fuzzing techniques to generate diverse inputs for contract execution, aiming to trigger hidden bugs in smart contracts.
Verx~\cite{permenev2020verx} and VeriSmart~\cite{so2020verismart} prove functional security properties for smart contracts based on formal verification techniques.
Securify~\cite{tsankov2018securify} and Ethainter~\cite{brent2020ethainter} use datalog analysis to conduct smart contact vulnerability detection.
Furthermore, researchers have explored integrating AI techniques with traditional program analysis methods to enhance the effectiveness of smart contract vulnerability detection~\cite{he2019learning,so2021smartest}.

\noindent
\textbf{Main Difference.}
Existing studies fail to detect finalization failure bugs for three reasons.
First, they cannot conduct tests for the finalization process on layer 2 protocols. Instead, they typically focus on the security of layer 1 blockchain (e.g., consensus protocols)~\cite{he2024nurgle,wu2024following,yang2021finding,chen2023tyr,luo2024towards}, or propose attacks against design flaws of layer 2 protocols~\cite{koegl2023attacks,medium_offchainlabs_rollup_delay}.
Second, they do not account for the unique design of zero-knowledge L2 protocols.
For example, despite generating transactions that may potentially reveal hidden bugs in Polygon zkRollup, existing tools~\cite{chen2023tyr} are unsuccessful in doing so. 
This is because these transactions will be discarded by the transaction pool during pre-execution phases due to their execution errors.
Third, existing studies rely on external factors, such as network partitioning~\cite{chen2023tyr,ma2023loki}, chaos testing~\cite{ma2023phoenix}, and node offline scenarios~\cite{chen2023tyr,ma2023loki,ma2023phoenix,su2023hybrid}, to conduct their tests. However, as Polygon zkRollup operates in a centralized manner, and is controlled by a permissioned entity, their testing methodologies are not applicable in the practical deployment environment of Polygon zkRollup.

\section{Conclusion}
Finalization failure bugs in zero-knowledge layer 2 protocols impair the usability of these protocols and the scalability of the main blockchains.
We conduct the first systematic study on finalization failure bugs in zero-knowledge layer 2 protocols, and define two kinds of such bugs.
Besides, we design \tool{}, the first tool to detect finalization failure bugs in Polygon zkRollup.
Our experimental results show that \tool{} can effectively detect twelve zero-day finalization failure bugs in Polygon zkRollup, significantly outperforming all baseline tools in both bug-finding capability and coverage.
Beyond bug detection, we derive key insights into why these 
bugs occur and how they can be fixed. 
These insights will not only 
enhance the security of Polygon zkRollup but also provide valuable guidance for other zero-knowledge layer 2 protocols.

\section*{Acknowledgements}
The authors thank the anonymous reviewers for their constructive comments.
This work is partly supported by Hong Kong RGC Projects (PolyU15224121 and PolyU15231223), National Natural Science Foundation of China (No.62172301 and No.62332004), and Sichuan Provincial Natural Science Foundation for Distinguished Young Scholars (No.2023NSFSC1963).

\bibliographystyle{ACM-Reference-Format}
\balance
\bibliography{ref}

\newpage
\appendix
\balance

\section{Transaction execution errors within Polygon zkRollup}
\label{sec_app_errors}

There are three kinds of execution errors in different scenarios within Polygon zkRollup, leading to distinct execution results.

\noindent
$\bullet$
i) Transaction execution errors are handled by error-handling opcodes or mechanisms within \rom{}. For example, opcodes like \zkcode{REVERT} are utilized to raise errors, reverting state modifications if the transaction execution or states fail to satisfy high-level business logic in contracts. Additionally, mechanisms are in place to handle errors such as stack underflow and jumping destination checking failures in \rom{}. In summary, \executor{} can handle these errors, terminating the transaction execution. During the proof generation process, these errors do not disrupt the process.

\noindent
$\bullet$
ii) Out-of-counter errors. As mentioned in~\S\ref{sec_back_archi}, transaction execution is measured by counters. If the consumed counters exceed predefined thresholds, \executor{} terminates the transaction execution to prevent unlimited resource consumption in generating the corresponding zk proof. Despite the premature termination, zk proofs for these transactions can still be generated during the proof generation process. However, during the pre-execution phase within Txpool, transactions triggering such errors are discarded~\cite{polygon_zkevm_doc}.

\noindent
$\bullet$
iii) Unexpected errors refer to errors that are not handled by \executor{}. These errors include cases where constraints are not satisfied during execution within \rom{}, or when the accessed memory region exceeds the valid maximum memory region. In such cases, \executor{} terminates the transaction execution. However, generating a proof for the transaction execution becomes impossible due to these unexpected errors, which fall beyond the assumptions of proof schemes, e.g., the accessed memory locations must be limited to 0x20000.
Besides, during the pre-execution phase within Txpool, transactions triggering such errors will also be discarded.

\section{Case studies for detected bugs on Polygon zkRollup}
\label{sec_bug_polygon}

\noindent
\textit{-- Bug \#1.} 
During the sanity check for a batch, \sequencer{} will forward the batch to \executor{} to check for any errors while reprocessing this batch.
When this bug triggers,
\sequencer{} halts from producing subsequent batches and L2 blocks and enters an infinite loop to continuously report errors, when the allocated execution step limit (i.e., counters~\cite{polygon_zkevm_doc}) for \executor{} is surpassed during executing a batch.
The root cause for this issue is that the maximum execution steps allowed for a batch to be executed by \executor{} are smaller than the maximum execution steps required for \sequencer{} to assemble a batch.
Hence, when \sequencer{} forwards a batch whose execution steps exceed \executor{}'s maximum limit, \executor{} triggers an out-of-counter error. 
This error finally results in \sequencer{} being trapped to continuously report errors while processing the batch as part of sanity check for this batch~\cite{wood2014ethereum}.

\noindent
\textit{-- Bug \#2.} 
\sequencer{} becomes incapable of proposing further batches and L2 blocks as it is compelled to continuously reprocess several high-efficiency transactions, each of which exceeds the execution step limit of \sequencer{} for assembling a batch from the transaction. 
The root cause is that \executor{} can execute and identify a transaction as valid even when its execution steps exceed those required for \sequencer{} to assemble a batch. 
Therefore, after \executor{} executes a transaction and identifies that its execution steps exceed \sequencer{}'s limit for batch assembly, \sequencer{} will exclude the transaction from the current batch and return it back to the txpool for inclusion in future batches. These high-efficiency transactions remain in the txpool of \sequencer{}, continuously causing \sequencer{} to prioritize reprocessing them, which prevents \sequencer{} from proposing new batches and L2 blocks.

\noindent
\textit{-- Bug \#3.} 
\sequencer{} becomes unable to propose new batches and L2 blocks because it is forced to repeatedly process several high-efficiency transactions.
This issue arises from \sequencer{}'s incorrect computation of transaction senders (e.g., $\textit{sender}_{\textit{seq}}$), which differ from those computed by \executor{} (e.g., $\textit{sender}_{\textit{exe}}$). 
This inconsistency can lead to a situation where (i) \executor{} deems the transaction senders $\textit{sender}_{\textit{exe}}$ to have insufficient funds, thus returning the transactions to \sequencer{} without incurring any state transition and any cost for the senders, and (ii) \sequencer{} considers the senders $\textit{sender}_{\textit{seq}}$ as having sufficient funds, and reserves the transactions in its txpool for further processing. 
The underlying error in \sequencer{}'s incorrect calculations for transaction senders stems from \sequencer{}'s incorrect method of determining whether a transaction is compatible to EIP-155~\cite{buterin2016eip155}. 
Therefore, when \sequencer{} incorrectly determines a transaction as EIP-155 incompatible, it employs wrong methods to derive the transaction sender for this transaction, leading to the inconsistency with \executor{}'s calculations for the transaction sender. 
As a result,
high-efficiency transactions that trigger the above situation are reserved in \sequencer{}'s txpool, continuously compelling \sequencer{} to prioritize them to execute by Executor,  preventing \sequencer{} from proposing new batches and L2 blocks.

\noindent
\textit{-- Bug \#4.} During the procedure for conducting long division operations, \rom{} will copy data from the array \zkcode{array\_mul\_out} to the array \zkcode{array\_add\_AGTB\_inA}. However, the maximum length of the former array \zkcode{array\_mul\_out} (i.e., \zkcode{\%ARRAY\_MAX\_LEN\_DOUBLED}) exceeds the maximum length of latter array \zkcode{array\_add\_AGTB\_inA} (i.e., \zkcode{\%ARRAY\_MAX\_LEN}).
Consequently, when the actual length of the former array \zkcode{array\_mul\_out} exceeds the maximum length of latter array (i.e., \zkcode{\%ARRAY\_MAX\_LEN}), data copying during long division operations will lead to overflow.
This overflow will result in corresponding constraints being unsatisfied, thereby causing the failure in the proof generation process for transaction execution.
In cases where a transaction within a forced batch triggers such an issue during the long division operations, this transaction will force \executor{} to proceed
with its execution despite its inability to generate a valid proof. As a result, this incapacity prevents the generation of valid proofs for the execution of all subsequent batches.

\noindent
\textit{-- Bug \#5.} 
\sequencer{} becomes unable to propose new batches and L2 blocks because it is forced to repeatedly process several high-efficiency transactions. This issue arises because \sequencer{} mishandles the error \zkcode{ZKR\_SM\_MAIN\_ADDRESS} raised by \executor{}. Specifically, \sequencer{} continues to regard transactions triggering this error as valid, reserving them in the txpool without incurring any state transition and any cost to the senders.
The error \zkcode{ZKR\_SM\_MAIN\_ADDRESS} occurs when the maximum memory offset accessed during transaction execution exceeds the memory offset set by \executor{} (e.g., \zkcode{0x10000}).
Consequently, high-efficiency transactions that trigger
this error are reserved in Sequencer’s txpool, continuously compelling Sequencer to prioritize them to execute by Executor, preventing Sequencer from proposing new batches and L2 blocks.

\noindent
\textit{-- Bug \#6.}
\executor{} halts the proof generation for a transaction's execution due to an unexpected error \zkcode{ZKR\_SM\_MAIN\_OOC\_MEM\_ALIGN}. This error occurs when the count of memory align operations, i.e., reading and writing a 32-byte word to memory, exceeds the permissible limit.
This issue stems from \executor{} inaccurately counting the number of memory align operations during a transaction's execution.
Specifically, when executing the \zkcode{CALLDATACOPY} opcode, \executor{} will repetitively copy 32 bytes of data from the input field of the transaction~\cite{wood2014ethereum} into memory,  involving memory align operations.
However, \executor{} only checks the count of available memory align operations before beginning the repetitive data copy, and tallies the performed memory align operations at once. Therefore, during this repetitive data copy process, \executor{} fails to recognize when the limit for memory align operations is surpassed, leading to the unexpected error.
In cases where a transaction within a forced batch triggers such an unexpected issue during execution, this transaction will force \executor{} to proceed with its execution despite its inability to generate a valid proof. As a result, this incapacity prevents the generation of valid proofs for the execution of all subsequent batches.

\noindent
\textit{-- Bug \#7.} 
During the underlying procedure of finite field arithmetic computation involved in proof generation process (e.g., witness generation), \executor{} neglects to deallocate the initialized finite field variables. This oversight issue results in a memory leak error, leading to the failure in the proof generation for corresponding transactions' execution. In cases where transactions within a forced batch induce this memory leak error during the proof generation process, these transactions will lead \executor{} to proceed with their execution despite its inability to generate a valid proof. Consequently, this incapacity prevents the generation of valid proofs for the execution of all subsequent batches.

\noindent
\textit{-- Bug \#8.}
\executor{} throws an exception and halts the process of proof generation when the last 2048 bytes of the valid region of memory have been accessed. This issue arises from an incorrect implementation in \executor{}, where the last 2048 bytes of the valid region of memory are set as inaccessible. In cases where a transaction within a forced batch triggers such an unexpected issue during execution, this transaction
will force \executor{} to proceed with its execution despite its inability to generate a valid proof. Consequently, this incapacity prevents the generation of valid proofs for the execution of all subsequent batches.

\noindent
\textit{-- Bug \#9.} 
During the procedure for converting data formats of incoming transactions' metadata, \sequencer{} is halted from producing subsequent batches and L2 blocks, and enters an infinite loop, continuously reporting errors when \sequencer{} receives a transaction with an odd length for the \zkcode{effectivePercentage} variable.
In such instances, an error of \zkcode{encoding/hex: odd length hex string} arises during the conversion of the \zkcode{effectivePercentage} variable of the transaction to the \zkcode{[]byte} type. This error subsequently results in \sequencer{} being trapped in an endless loop, continuously reporting errors while processing the transaction.

\noindent
\textit{-- Bug \#10.} 
During the process of appending a new transaction to the txpool, the mutex \zkcode{workerMutex} in \sequencer{} will be double unlocked if the state database is inaccessible for \sequencer{}, resulting in the crash of \sequencer{}.
Specifically, \sequencer{} acquires a mutex on \zkcode{workerMutex} at the beginning of the process for adding the transaction to the txpool.
Besides, \sequencer{} schedules to automatically unlock \zkcode{workerMutex} upon termination of the process by using the keyword \zkcode{defer}. 
During this process, \sequencer{} temporarily unlocks \zkcode{workerMutex} to perform additional checks on the transaction sender (e.g., checking the sender's available funds) using information from state database. 
However, if the state database become temporarily inaccessible for \sequencer{} due to exceeding the maximum connection limit, the process will terminate prematurely leaving \zkcode{workerMutex} unlocked.  
In conjunction with the mutex unlocking instructions triggered by \zkcode{defer}, \zkcode{workerMutex} will be unlocked twice, thereby leading to the crash of \sequencer{}.

\noindent
\textit{-- Bug \#11.} 
\executor{} encounters errors and halts the proof generation process for a transaction's execution when the accessed maximum memory offset exceeds the checked boundary. This issue stems from \executor{}'s incorrect implementation of memory boundary checks, i.e., incorrectly flagging normal memory offsets as exceeding the valid memory boundary.
In cases where a transaction within a forced batch triggers such an unexpected issue during execution, this transaction
will force \executor{} to proceed with its execution despite its
inability to generate a valid proof. As a result, this incapacity
prevents the generation of valid proofs for the execution of
all subsequent batches.

\noindent
\textit{-- Bug \#12.} 
The long division operation within the modexp procedure~\cite{polygon_zkevm_doc} does not account for the scenario of integer division, where the remainder is 0 when the dividend is divisible by the divisor. Hence, when a integer division occurs, resulting in a remainder of 0, the corresponding constraints for the remainder cannot be satisfied (e.g., the remainder must be greater than or equal to 1).
The unsatisfied constraints for the reminder further cause the proof generation process for transaction execution halts.
In cases where a transaction within a forced batch triggers such an issue during the long division operation within modexp, this transaction will force \executor{} to proceed with its execution despite its inability to generate a valid proof.
As a result, this incapacity prevents the generation of valid proofs for the execution of all subsequent batches.

\section{Coverage trends of evaluated tools in RQ2}
\label{sec_comparison_coverage}
\begin{figure}[b]
	\centering
	\includegraphics[width=0.99\linewidth]{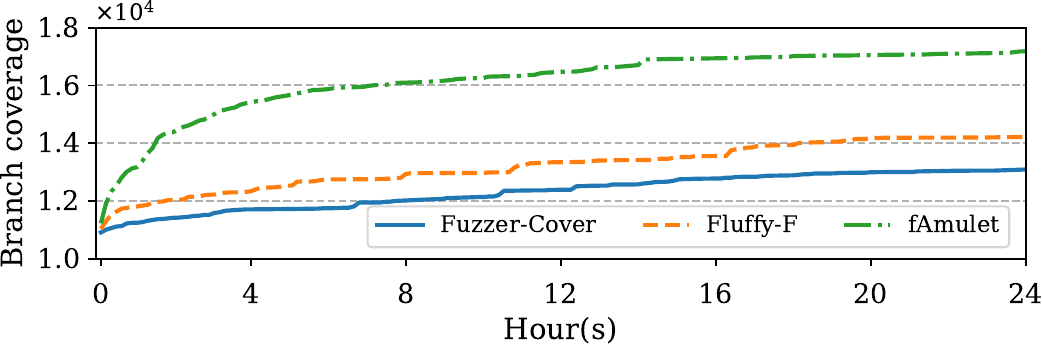}
	\caption{Trends of coverage growth over time for each tool}
	\label{fig_coverage}
\end{figure}
Fig.\ref{fig_coverage} shows that all tools' coverage significantly increases within the first hour. 
Subsequently, \tool{} consistently covers more branches, whereas the coverage of baselines gradually converges.

\section{Cases studies for detected bugs on other L2s}
\label{sec_other_bugs}

-- The zero-day bug detected by us in Scroll zkRollup causes \sequencer{} to be unable to accept new user transactions, thereby preventing it from generating subsequent batches and L2 blocks.
This bug arises due to unavailable RPC services resulting from the memory resource exhaustion of RPC nodes.
Specifically, RPC nodes fail to set an upper limit on the number of requests that can be included in a batch request for RPC services~\cite{polygonrpcs}. Consequently, when receiving batch requests, the results of each request are stored in a list in the memory, and only when all requests in a batch request are responded can RPC nodes return these results to users.
Even worse, the entire process of preparing results for the requests in a batch request will produce multiple copies of all results in RPC nodes' memory (e.g., copies produced by JSON encoding). As a result, memory-consuming requests, such as those retrieving transactions with metadata at the maximal size, can efficiently exhaust RPC nodes' memory resources, triggering their crash.

\noindent
-- The first known bug detected by us in Optimism Rollup prevents \sequencer{} from accepting new transactions, thereby hindering the generation of subsequent batches and L2 blocks. This issue arises due to unavailable RPC services resulting from the crash of RPC nodes. Specifically, RPC nodes fail to check whether the \zkcode{dec.Gas} field of a deposit transaction is \zkcode{nil}. Hence, when receiving deposit transactions without the gas field, the nil pointer dereference of \zkcode{dec.Gas} triggers a runtime error, causing RPC nodes to crash.

\noindent
-- The second known bug detected by us in Optimism Rollup causes unexpected errors in \sequencer{}, leading to its execution halt and preventing the production of subsequent batches and L2 blocks. This issue arises during the procedure for converting transactions in an L2 block into a batch transaction to be submitted to the L1 blockchain. In this process, \sequencer{} assumes the first transaction in the block is a deposit transaction, and unmarshals the corresponding meta information of the deposit operations from the transaction's data payload by using \zkcode{L1InfoDepositTxData()}. However, \sequencer{} fails to verify the type of the first transaction in the L2 block before calling \zkcode{L1InfoDepositTxData()}. Consequently, when processing an L2 block with a non-deposit transaction as the first transaction, \sequencer{} encounters unexpected errors while handling the invalid data payload, leading to execution failure.

\end{document}